\documentclass[11pt, oneside]{article}

\usepackage[english]{babel}
\usepackage[utf8]{inputenc}
\usepackage[T1]{fontenc}

\usepackage[letterpaper,top=1in,bottom=1in,left=1in,right=1in,marginparwidth=1in]{geometry}

\usepackage{amsmath}
\usepackage{graphicx}
\graphicspath{{./figs/}{./}}
\usepackage{url}
\usepackage[colorinlistoftodos]{todonotes}
\usepackage[colorlinks=true, allcolors=blue]{hyperref}
\usepackage{listings}                   
\usepackage{float}
\usepackage{mathtools}
\usepackage{braket}
\usepackage{slashed}
\usepackage[bottom]{footmisc}
\usepackage{pdfpages}
\usepackage{amsfonts,amsthm,amssymb}
\usepackage{type1cm}
\usepackage{lettrine}
\usepackage{sectsty}
\usepackage{titlesec}
\usepackage{cleveref}
\numberwithin{equation}{section}
\usepackage{setspace}
\usepackage[parfill]{parskip}
\usepackage{cite}
\usepackage[margin=2em]{caption}

\title{\textbf{Conservation laws and effective hadronization models}}

\author{Tony Menzo$^{1,2}$\footnote{\texttt{amenzo@ua.edu}} \vspace{0.2in} \\  
${^1}$\textit{Department of Physics and Astronomy, University of Alabama, Tuscaloosa, AL 35487, USA}\\
${^2}$\textit{Theoretical Physics Department, Fermilab, Batavia, IL 60510, USA}
}

\begin{document}
\pagenumbering{gobble}
\maketitle

\vspace{0.1in}

\begin{center}
\begin{minipage}{14cm}

\begin{center}
\textbf{\Large Abstract}
\end{center}
\vspace{0.5em}

Hadronization models based on local string-breaking dynamics are typically Markovian by construction, yet the physical ensemble of final states is shaped by global constraints that couple the entire fragmentation trajectory.
Recasting hadronization as a conditioned stochastic diffusion process provides a precise mathematical resolution to this tension.
In particular, this language reveals explicitly that constraints stemming from conservation laws induce non-Markovian correlations between otherwise independent fragmentation steps, and that these correlations can be absorbed exactly into a renormalization of the local dynamics through a Doob $h$-transform.
We develop this formalism for a $q\bar{q}$ string in the chiral limit, where the longitudinal-transverse factorization of the Lund kernel becomes exact, enabling systematic power counting and clean ultraviolet (UV)/infrared (IR) separation.
The dynamics organize naturally into a tower of effective theories distinguished by the remaining string mass, spanning a UV fixed point with scale-invariant transport coefficients, an intermediate regime where transverse phase space induces controlled running, and an IR boundary layer where non-local effects enter at leading order.
The tower exhibits genuine Wilsonian structure, including $\beta$-functions, anomalous dimensions, and systematic matching conditions.
The resulting framework achieves a clean factorization of universal microscopic fragmentation dynamics from infrared constraint effects, and opens new directions for both the theoretical analysis and practical simulation of hadronization.
\end{minipage}
\end{center}
\vspace*{1cm}

\newpage
\tableofcontents
\newpage

\pagenumbering{arabic}
\setcounter{page}{1}

\section{Introduction}
\label{sec:introduction}

Hadronization, the process by which asymptotically free quarks and gluons form bound hadronic states, remains one of the principal sources of theoretical uncertainty across particle physics. 
The accuracy of hadronization modeling directly impacts extractions of the strong coupling constant from event shapes and jet rates~\cite{dEnterria:2022hzv,Kardos:2018kqj,Verbytskyi:2019zhh}, determinations of the top quark mass~\cite{Hoang:2020iah}, measurements of the $W$ boson mass via direct reconstruction~\cite{Bozzi:2025xtr,ALEPH:2013dgf}, jet energy scale calibrations~\cite{ATLAS:2020cli} and substructure observables~\cite{Buckley:2011ms}, precision electroweak pseudo-observables at future Higgs factories~\cite{Peskin:2025LP}, and neutrino energy reconstruction in long-baseline oscillation experiments~\cite{Katori:2015glr}. 
Despite decades of phenomenological success, Monte Carlo event generators rely on largely heuristic models, and there exists no systematically improvable framework connecting microscopic hadronization dynamics to macroscopic hadronic observables.

Phenomenological hadronization models fall into two broad classes: string-based~\cite{Andersson:1997xwk, Andersson:1983ia, Sjostrand1984469}, on which this work builds, and cluster-based~\cite{Webber:1984kz}. 
The Lund string model, implemented in \textsc{Pythia}~\cite{Sjostrand:1982fn,Sjostrand:2006za,Sjostrand:2014zea}, models the color flux tube between separating partons as a relativistic string that fragments through iterative, causally disconnected $q'\bar{q}'$ pair production from the vacuum.
The breakup probability is governed by the invariant worldsheet area swept out between adjacent vertices, and the longitudinal splitting function is uniquely fixed by left-right symmetry~\cite{Andersson:1983ia}.
The cluster model, implemented in \textsc{Herwig}~\cite{Corcella:2000bw,Bahr:2008pv,Bellm:2015jjp} and \textsc{Sherpa}%
\footnote{\textsc{Sherpa} implements both 
cluster-based hadronization (AHADIC++~\cite{Winter:2003tt}), and provides an interface to \textsc{Pythia}'s (\texttt{v6.4}) \cite{Sjostrand:2006za} Lund string fragmentation for uncertainty evaluation.}~\cite{Sherpa:2019gpd}, exploits the preconfinement property of perturbative QCD, splitting gluons nonperturbatively into quark-antiquark pairs whose color-singlet clusters decay isotropically according to phase space. 
Both models have been extensively tuned to describe a wide range of collider data~\cite{Skands:2010ak}. 

At the microscopic level string fragmentation is local and Markovian: each $q'\bar{q}'$ pair is produced independently, conserving energy-momentum and flavor at each vertex, such that the measure over complete fragmentation histories factorizes into a product over individual breaks~\cite{Andersson:1983ia,Andersson:1997xwk}.
Physical hadronization, however, must also satisfy global constraints---most fundamentally exact energy-momentum conservation---that act on the entire fragmentation trajectory.
In practical implementations such as \textsc{Pythia}, this tension is resolved through a joining procedure in which the string is fragmented iteratively from both endpoints until a low-mass remnant remains, which is then forced into a two-body decay to close the kinematics~\cite{Sjostrand:2006za,Sjostrand:2014zea}.
While these endpoint corrections have negligible impact on inclusive observables at high energies, they introduce systematic deviations from the underlying area law that become significant for few-body final states, moderate energies, and differential observables sensitive to rank-rapidity correlations~\cite{Andersson:1999ui,Eden:2000in}.
Alternative approaches based on rejection sampling over the complete area law~\cite{Andersson:1999ui,Eden:2000in} restore theoretical consistency but at computational cost that scales poorly with multiplicity.
The approaches developed to date either sacrifice theoretical rigor 
for computational efficiency, or achieve exactness at prohibitive cost, 
leaving open the question of whether a systematic and practical 
framework can reconcile local dynamics with global conservation laws.

The central insight of this work is to recast hadronization as a stochastic process governing the evolution of the string mass. 
This reformulation enables a systematic treatment of global constraints, \textit{i.e.}~conservation laws, and their influence on the probability measure over physical fragmentation trajectories.
In particular, it makes explicit that imposing conservation laws on the space of trajectories generically induces non-Markovian dynamics, and provides a natural framework for restoring Markovianity by absorbing these effects into a renormalization of the local dynamics.
This renormalization can be implemented exactly through a so-called Doob $h$-transform~\cite{Doob:1957,RogersWilliams:2000}, which reweights local transition probabilities according to the future viability of the cascade. 
By computing the success probability \textit{i.e.}~the probability that a fragmentation initiated at a given string mass terminates in a viable hadronic final state, we show that its gradient induces a multiplicative correction to the bare drift coefficient, biasing the evolution away from fragmentation trajectories that cannot be completed consistently. 
The interpretation of string fragmentation as a stochastic process was recognized in earlier work~\cite{Andersson:1985qr, Andersson:1993mv}, which identified connections to Brownian motion in the asymptotic regime where boundary effects are negligible. 
The present work develops the full machinery of conditioned processes necessary when global constraints cannot be ignored.

This reframing also reveals that the dynamics of string-hadronization naturally organize into a tower of \textit{effective hadronization models} distinguished by the remaining string mass and characterized by a systematic power counting. 
This tower exhibits many of the characteristic features of a Wilsonian effective field theory, including a scale-invariant ultraviolet fixed point, intermediate evolution governed by transverse phase space, systematic matching at threshold scales, and the emergence of explicit non-local structures associated with rare events. 
The observation that hadronization admits a well-defined renormalization group interpretation provides a transparent hierarchy of approximations with controlled accuracy across kinematic regimes. 
The resulting framework factorizes the full hadronization process into universal UV dynamics and constraint-dependent IR effects, and the localization of conditioning effects to the infrared boundary layer informs a hybrid sampling scheme that eliminates the rejection overhead of conventional approaches.

To develop the formalism in its cleanest form, we restrict attention throughout to a single $q\bar{q}$ string fragmenting into pions in the chiral limit $m_\pi \to 0$, where the Lund kernel exhibits exact longitudinal-transverse factorization. 
These simplifications isolate the structural ingredients of the EFT construction while deferring extensions to finite hadron masses, gluonic strings, and realistic termination conditions to future work.

The remainder of this paper is organized as follows. 
In Sec.~\ref{sec:stoch_dynamics}, we review the general theory of conditioned stochastic processes and introduce the Doob $h$-transform, establishing the mathematical framework for restoring Markovianity under global constraints. 
In Sec.~\ref{sec:hadronization}, we apply this formalism to string fragmentation, deriving the bare fragmentation kernel and the evolution equation for the string mass. 
Sec.~\ref{sec:effective_had_models} develops the effective theory tower for hadronization, identifying distinct kinematic regimes and establishing power counting rules. 
In Sec.~\ref{sec:h_global}, we compute the success probability in each effective theory and derive matching conditions. 
Sec.~\ref{sec:doob_h} presents the renormalized fragmentation dynamics via the $h$-transform and discusses the physical interpretation of drift renormalization as an emergent force. 
Sec.~\ref{sec:practical} analyzes the computational cost of conditioned sampling and introduces a hybrid scheme that applies the $h$-transform selectively in the boundary layer.
We conclude in Sec.~\ref{sec:conclusions} with a summary of results, implications for hadronization modeling, and avenues for future work. 
Supporting derivations are collected in the appendices: App.~\ref{app:moments} derives the moment expansion of the fragmentation kernel, App.~\ref{app:tail_operators} evaluates the non-local tail operators, and App.~\ref{app:constrained_dynamics} develops a pedagogical toy model of constrained Brownian motion illustrating the Doob $h$-transform.

\section{Stochastic dynamics and conservation laws}\label{sec:stoch_dynamics}

As discussed in the previous section, at the microscopic level, string fragmentation is governed by a local string-breaking law in which each break is independent and memoryless.  
The corresponding probability measure over complete fragmentation histories factorizes into a product over individual breaks. 
We refer to this factorized measure as the \emph{bare} or \emph{ultraviolet} dynamics. 
By contrast, global constraints such as energy-momentum conservation act on the fully realized set of hadron momenta, correlating breaks that would otherwise be independent. 
The physical hadronization measure is obtained by conditioning the bare dynamics on these global constraints. 
In this section, we show how this conditioning can be implemented while preserving a Markovian description at the level of local transition probabilities.

\subsection{Discrete dynamics and the Doob transform}

Consider an evolution variable $t$ that parametrizes the sequential construction of a fragmentation history, such as the running string mass $M$. Let $X_t$ denote the instantaneous state of the remaining string at ``time'' $t$.
At the level of the \emph{bare} fragmentation dynamics, the evolution of $X_t$ is Markovian. The bare probability measure $\mathbb{P}_{\text{bare}}$ induced by local string breaking prior to the imposition of global constraints is encoded by a one-step transition kernel
\begin{equation}
  \mathcal{K}_{\text{bare}}(x \to y; \Delta t)
  \;\equiv\;
  \mathbb{P}_{\text{bare}}\!\left(X_{t+\Delta t}=y \,\big|\, X_t=x \right),
  \label{eq:markov_kernel}
\end{equation}
which depends only on the current state $x$ of the string. The Markovian property ensures that the probability for a sequence of transitions factorizes as
\begin{equation}
  \mathbb{P}_{\text{bare}}\!\left(
    X_{t_1}=x_1,\ldots,X_{t_n}=x_n \,\big|\, X_{t_0}=x_0
  \right)
  \;=\;
  \prod_{i=0}^{n-1}
  \mathcal{K}_{\text{bare}}(x_i \to x_{i+1}; \Delta t_i).
  \label{eq:markov_factorization}
\end{equation}
The kernel $\mathcal{K}_{\text{bare}}$ governs the evolution of expectation values for any observable $f(x)$ defined on the state space. Given the current state $x$, the expected value of $f$ after one step is
\begin{equation}
  \mathbb{E}_{\text{bare}}[f(X_{t+\Delta t}) \mid X_t = x]
  =
  \int \mathrm{d}y\; \mathcal{K}_{\text{bare}}(x \to y; \Delta t)\, f(y).
  \label{eq:observable_evolution_discrete}
\end{equation}
This provides a complete characterization of the one-step dynamics: any statistical property of the evolution can be computed by choosing an appropriate observable $f$ and evaluating its conditional expectation.

Physical hadronization, however, is not governed by the bare kernel alone. 
The physically relevant probability measure is obtained by conditioning on a global constraint $\mathcal{C}$, such as exact energy-momentum conservation,
\begin{equation}
  \mathbb{P}_{\text{phys}}(\cdot)
  \;=\;
  \mathbb{P}_{\text{bare}}(\cdot \mid \mathcal{C}).
\end{equation}
Since $\mathcal{C}$ acts on complete fragmentation histories, its direct implementation would appear to induce non-local correlations along the trajectory. 
Consider the conditioned probability for a single transition $x \to y$. Applying Bayes' rule yields
\begin{equation}
  \mathbb{P}_{\text{phys}}\!\left(
    X_{t+\Delta t}=y \,\big|\, X_t=x, \mathcal{C}
  \right)
  =
  \frac{
    \mathbb{P}_{\text{bare}}\!\left(
      \mathcal{C} \,\big|\, X_{t+\Delta t}=y,\, X_t=x
    \right)
  }{
    \mathbb{P}_{\text{bare}}\!\left(
      \mathcal{C} \,\big|\, X_t=x
    \right)
  } \mathbb{P}_{\text{bare}}\!\left(
      X_{t+\Delta t}=y \,\big|\, X_t=x
    \right)
  \label{eq:bayes_step}
\end{equation}
where $\mathbb{P}_{\text{bare}}(\mathcal{C}\,|\,\cdot)$ denotes the conditional probability of satisfying the constraint given the current (or future) state. 
Using the Markov property of the bare dynamics, the probability of eventually satisfying the constraint $\mathcal{C}$ depends only on the current state of the system, so that
\begin{equation}
  \mathbb{P}_{\text{bare}}\!\left(
    \mathcal{C} \,\big|\, X_{t+\Delta t}=y,\, X_t=x
  \right)
  =
  \mathbb{P}_{\text{bare}}\!\left(
    \mathcal{C} \,\big|\, X_{t+\Delta t}=y
  \right).
\end{equation}
Defining the success probability
\begin{equation}
  h(x) \;\equiv\; \mathbb{P}_{\text{bare}}\!\left( \mathcal{C} \,\big|\, X_t=x \right),
  \label{eq:h_def}
\end{equation}
the conditioned dynamics can again be written in Markovian form, but with a \emph{renormalized} transition kernel,
\begin{equation}
  \mathcal{K}_h(x \to y; \Delta t)
  \;=\;
  \mathcal{K}_{\text{bare}}(x \to y; \Delta t)\,
  \frac{h(y)}{h(x)}.
  \label{eq:doobkernel_intro}
\end{equation}
This is the Doob $h$-transform. It makes explicit how a constraint imposed only on the final state renormalizes the local fragmentation dynamics at every intermediate step. 
The conditioned evolution of observables is then given by
\begin{equation}
  \mathbb{E}_{\text{phys}}[f(X_{t+\Delta t}) \mid X_t = x]
  =
  \int \mathrm{d}y\; \mathcal{K}_h(x \to y; \Delta t)\, f(y)
  =
  \frac{1}{h(x)}\int \mathrm{d}y\; \mathcal{K}_{\text{bare}}(x \to y; \Delta t)\, h(y)\, f(y),
  \label{eq:observable_conditioned_discrete}
\end{equation}
where the factor $h(y)/h(x)$ biases each transition according to the future viability of the cascade. The observable $f$ itself is unchanged; conditioning acts entirely through the renormalized kernel $\mathcal{K}_h$. Consequently,
\begin{equation}
  \mathbb{E}_{\text{phys}}[f] = \mathbb{E}_{\text{bare}}[f \mid \mathcal{C}],
  \label{eq:expectation_equivalence}
\end{equation}
so that the $h$-transform provides an exact means of computing conditional expectations without explicit rejection sampling. 
Crucially, because the bare dynamics are Markovian, the success probability $h(x)$ depends only on the current state, not on the prior history. 
The renormalized kernel $\mathcal{K}_h$ therefore remains Markovian and global constraints are absorbed into a local, state-dependent reweighting of transition probabilities.

\subsection{The continuum limit and generator renormalization}
\label{subsec:continuum_limit}

The discrete transition kernel $\mathcal{K}_{\text{bare}}$ provides an exact microscopic description of string fragmentation prior to the imposition of global constraints. 
While sufficient to define the dynamics at all scales, this formulation is not well adapted to analytic control or to the systematic organization of physical effects across different kinematic regimes. 
The scale dependence of observables and the interplay between local dynamics and global constraints become explicit once the continuum limit is taken. 
In this limit, the discrete kernel is replaced by an infinitesimal generator $\mathcal{L}$, defined as
\begin{equation}
  \mathcal{L} f(x) 
  \;=\; 
  \lim_{\Delta t \to 0} \frac{1}{\Delta t} 
  \left( 
    \mathbb{E}_{\text{bare}}[f(X_{t+\Delta t}) \mid X_t = x] - f(x) 
  \right),
  \label{eq:generator_def}
\end{equation} 
and encoding the instantaneous rate of change of observables under the bare dynamics.

When the remaining string mass is large compared to hadronic scales, fragmentation proceeds through many steps in which each individual transition induces only a small change in the state variable $x$. 
In this regime, the limit in \cref{eq:generator_def} exists and admits a controlled expansion. 
Substituting the kernel representation of the conditional expectation,
\begin{equation}
  \mathcal{L} f(x) 
  \;=\; 
  \lim_{\Delta t \to 0} \frac{1}{\Delta t} 
  \int \mathrm{d}y\; \mathcal{K}_{\text{bare}}(x \to y; \Delta t) 
  \left[ f(y) - f(x) \right],
\end{equation}
and expanding $f(y) = f(x + \Delta x)$ in moments of the step size $\Delta x \equiv y - x$ yields the Kramers--Moyal representation
\begin{equation}
  \mathcal{L}
  \;=\;
  \sum_{n=1}^{\infty} \frac{1}{n!}\,\kappa_n(x)\,\partial_x^n
  \;=\;
  \mu(x)\,\partial_x
  \;+\;
  \frac{1}{2}\,\sigma^2(x)\,\partial_x^2
  \;+\;\cdots,
  \label{eq:bare_generator}
\end{equation}
where the coefficients
\begin{equation}
  \kappa_1(x) = \langle \Delta x \rangle, \qquad
  \kappa_2(x) = \langle (\Delta x)^2 \rangle - \langle \Delta x \rangle^2, \qquad
  \kappa_3(x) = \langle (\Delta x)^3 \rangle_{\mathsf{c}},
  \label{eq:first_cumulants}
\end{equation}
are the first three cumulants of the step-size distribution induced by $\mathcal{K}_{\text{bare}}$ at state $x$, with $\langle \cdot \rangle$ denoting the conditional expectation over single transitions and higher cumulants defined analogously.
The first cumulant $\mu(x) \equiv \kappa_1(x)$ is the drift, encoding the mean displacement per step, and the second $\sigma^2(x) \equiv \kappa_2(x)$ is the diffusion coefficient, encoding the variance. When individual steps are small, higher cumulants scale as $\kappa_n \sim (\Delta x)^n$ and truncation at second order, \textit{a.k.a.}~the diffusion approximation, is controlled. 

The physical dynamics are obtained by conditioning the bare process on the global constraint $\mathcal{C}$. 
Just as the discrete kernel acquires a factor $h(y)/h(x)$ under conditioning, the continuum generator transforms by conjugation with $h$:
\begin{equation}
  \mathcal{L}_h
  \;\equiv\;
  h^{-1}\,\mathcal{L}\,h,
  \label{eq:doob_generator}
\end{equation}
which acts on observables through $\partial_t \mathbb{E}_{\text{phys}}[f] = \mathbb{E}_{\text{phys}}[\mathcal{L}_h f]$. 
This similarity transformation is the continuum Doob $h$-transform. Geometrically, $h(x)$ tilts the probability measure over trajectories, inducing an emergent drift proportional to $\partial_x \log h$ that biases paths toward regions of high success probability. 
A pedagogical illustration of the conditioning mechanism is developed with a toy model in appendix~\ref{app:constrained_dynamics}, where Brownian motion with drift subject to endpoint constraints is analyzed explicitly, demonstrating how the Doob $h$-transform induces an emergent force that biases trajectories toward successful termination.

From the perspective of trajectory space, conservation laws induce correlations between widely separated fragmentation steps, but these correlations can be mediated entirely through the local state-dependent factor $h(x)$. 
This is the stochastic analogue of integrating out forbidden field configurations in a Wilsonian effective theory. 
The ultraviolet kernel $\mathcal{K}_{\text{bare}}$ is local and factorized, while the physical kernel $\mathcal{K}_h$ incorporates infrared constraints through a scale-dependent renormalization. 
In Monte Carlo language, rejection and endpoint-correction procedures are empirical ways of implementing the reweighting in \cref{eq:doobkernel_intro}. 
With this framework in hand, we now turn to its application in hadronization.

\section{Hadronization}
\label{sec:hadronization}

At large distances, confinement in QCD implies that the interaction between a quark and an antiquark is well approximated by a linear potential~\cite{Bali:1992ab,Bali:1997am},
\begin{equation}
  V_{q\bar q}(r) \;\simeq\; \kappa\, r ,
\end{equation}
with string tension $\kappa \sim 1~\mathrm{GeV/fm} \simeq 0.2~\mathrm{GeV}^2$.
The Lund string model adopts this picture as an effective description of nonperturbative QCD dynamics, treating the color flux tube as a relativistic string with constant energy density that fragments through a sequence of causally disconnected, local $q'\bar{q}'$ pair creations from the vacuum.
Physical hadronization must also satisfy global conservation laws, including exact energy-momentum conservation, net flavor, and (approximate) isospin, none of which appear in the local string-breaking law. 
It is important to distinguish \emph{local} from \emph{global} conservation in this context.
Each individual string break is kinematically consistent, with the produced hadron and the remaining string together conserving energy-momentum at that step.
The global constraint, however, acts on the \emph{complete} fragmentation history.
The local Lund kernel stochastically samples hadron kinematics without reference to how much kinematic budget remains for subsequent breaks.
A sequence of individually valid local breaks may therefore fail globally by, for instance, leaving a remnant string with insufficient mass to produce even the lightest hadrons.
It is this tension between locally valid stochastic dynamics and globally consistent termination that the conditioned stochastic framework is designed to address.

Hadronization therefore presents a natural application of the constrained stochastic framework developed in \cref{sec:stoch_dynamics}.
In this work, we restrict attention to the simplest hadronizing system: a single $q\bar q$ string with light-flavor endpoints (up, down) in its center-of-mass frame, fragmenting exclusively into pions. 
This system contains all structural ingredients required for the effective construction developed below.
The generalization to include other hadron species (kaons, baryons, etc.) is conceptually straightforward, only requiring additional string end flavors and additional book-keeping.

\subsection{The bare fragmentation measure}

To make the connection to \cref{sec:stoch_dynamics} explicit, we begin by identifying the local and global structures in the Lund model.
The differential probability to fragment a string into an $n$-hadron final state may be written schematically as \cite{Andersson:1997xwk}
\begin{equation}
  \mathrm{d}P_n \;\propto\;
  \left[
    \prod_{i=1}^n \mathrm{d}^4 p_i \, \delta(p_i^2 - m_i^2)
  \right]
  \delta^{(4)}\!\left(P_{\rm tot} - \sum_{i=1}^n p_i \right)
  \;
  \prod_{k=1}^{n-1}
  \left[
    \kappa \, \mathrm{d}A_k \,
    \exp\!\left(-\frac{\pi m_{\perp,k}^2}{\kappa}\right)
  \right],
  \label{eq:dPn}
\end{equation}
where $P_{\rm tot}$ is the total string four-momentum, $\kappa$ is the string tension, $\mathrm{d}A_k$ is the invariant worldsheet area associated with the $k$-th string break, and $m_{\perp,k}^2 = m_k^2 + p_{\perp,k}^2$ is the transverse mass of the produced $q'\bar q'$ pair. 
This expression mixes two distinct structures: (i) a \emph{local} worldsheet intensity for producing breaks, which factorizes over independent string-breaking events, and (ii) a \emph{global} constraint on the \emph{entire} history, encoded by the final delta function.

\begin{figure}[t]
\centering
\begin{minipage}[c]{0.52\textwidth}
\centering
\includegraphics[width=\textwidth]{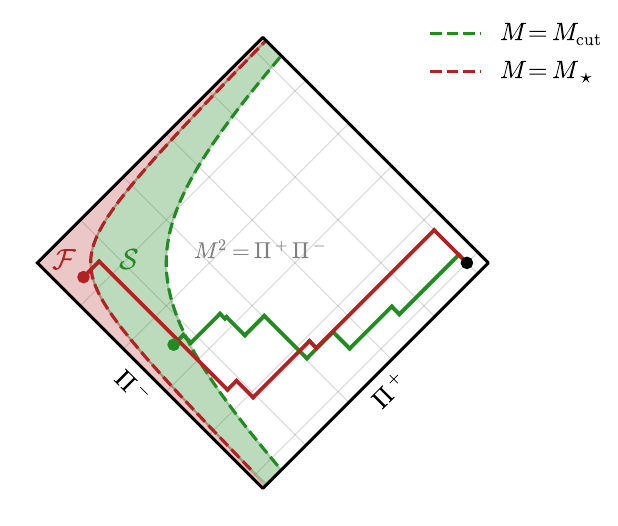}
\end{minipage}%
\begin{minipage}[c]{0.43\textwidth}
\centering
\includegraphics[width=\textwidth]{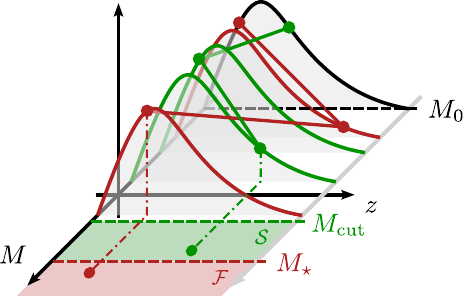}
\end{minipage}
\caption{Two complementary depictions of string hadronization. 
\textbf{Left:} The Lund plane representation showing string breaks in light-cone coordinates, where each hadron carries a fraction $z$ of the available light-cone momentum and transverse mass $m_\perp^2$. The worldsheet area element controls the probability for string breaking via tunneling. 
\textbf{Right:} The stochastic mass evolution representation, where hadronization is viewed as a discrete Markov chain governing the evolution of the remaining string mass $M_n \to M_{n+1}$, with each step corresponding to a single hadron emission. The cascade terminates when the mass enters the termination band $\mathcal{S} = [M_\star, M_{\rm cut}]$ or fails by undershooting into $\mathcal{F} = (0, M_\star)$.}
\label{fig:discrete_dynamics}
\end{figure}

The separation becomes transparent when we rewrite \cref{eq:dPn} as an unconstrained trajectory measure multiplied by a global constraint functional.
Let $\Pi^\pm$ denote the light-cone momenta of the remaining string, with string mass $M^2 = \Pi^+ \Pi^-$.
A complete fragmentation history $\omega \equiv \{(\Pi^\pm_k, p_k)\}_{k=0}^{n-1}$ specifies the string state and emitted hadron momentum at each step, as depicted in \cref{fig:discrete_dynamics}. 
Define the bare (ultraviolet) weight
\begin{equation}
  \mathrm{d}\mathbb{P}_{\text{bare}}(\omega)
  \;\propto\;
  \prod_{k=1}^{n-1}
    \left[ \kappa \, \mathrm{d}A_k \,
    \exp\!\left(-\frac{\pi m_{\perp,k}^2}{\kappa}\right) \right] \,
  \times \Big(\text{local phase space factors}\Big),
  \label{eq:P0}
\end{equation}
which factorizes over breaks by construction.
Then \cref{eq:dPn} becomes
\begin{equation}
  \mathrm{d}\mathbb{P}(\omega)
  \;\propto\;
  \mathrm{d}\mathbb{P}_{\text{bare}}(\omega)\;
  \delta^{(4)}\!\Big(P_{\rm tot}-\sum_{i=1}^n p_i(\omega)\Big).
  \label{eq:Pdelta}
\end{equation}

This is precisely the structure analyzed in \cref{sec:stoch_dynamics}: a factorized bare measure $\mathbb{P}_{\text{bare}}$ reweighted by a constraint functional that depends on the entire trajectory. 
The bare UV law generates histories without ``foresight''. 
Each string break is blind to the global kinematic budget, and conservation is enforced only after the history is complete.
For generic histories drawn from $\mathbb{P}_{\text{bare}}$, the constraint functional in \cref{eq:Pdelta} is essentially never satisfied.
This tension between local dynamics and global constraints is precisely the mechanism by which conditioning induces non-Markovianity.

\subsection{The bare fragmentation kernel}
\label{sec:lund_kernel}

In the Lund model, hadronization proceeds through an iterative sequence of string breaks.
At each step, a hadron of transverse mass $m_\perp$ is produced and absorbs a fraction $z$ of the longitudinal momentum of the remaining string.
A complete specification of the single-step fragmentation kernel therefore requires a distribution over the longitudinal momentum fraction $z$ together with a prescription for the transverse mass $m_\perp^2$.

Requiring that the fragmentation process admit a symmetric description whether constructed iteratively from the quark end or from the antiquark end (left-right symmetry), combined with Lorentz invariance and a central rapidity plateau at high energies, uniquely determines the longitudinal momentum distribution~\cite{Andersson:1983ia,Andersson:1983jt}.
For any fixed value of $m_\perp^2$, the conditional distribution over $z$ is
\begin{equation}
  f(z \,|\, m_\perp^2)\,\mathrm{d}z
  \;\propto\;
  \frac{(1-z)^a}{z}\,
  \exp\!\left( -b\,\frac{m_\perp^2}{z} \right)\mathrm{d}z,
  \label{eq:lund_ff_conditional}
\end{equation}
where $a$ and $b$ are phenomenological parameters. 
Crucially, the derivation treats $m_\perp^2$ as a parameter. 
It determines how longitudinal momentum is shared for a given transverse mass, but it does not specify how transverse masses themselves are distributed.
The marginal distribution $p(m_\perp^2)$ is therefore not fixed by left-right symmetry and must be supplied by additional modeling assumptions.

In practical implementations of string fragmentation, the transverse sector is commonly motivated by the Schwinger tunneling mechanism, which describes spontaneous pair production in a uniform electric field~\cite{Schwinger:1951nm,Casher:1978wy}. 
When applied heuristically to a chromoelectric flux tube, this picture yields a production rate for a $q'\bar q'$ pair with quark mass $m_q$ and transverse momentum $p_\perp$ proportional to $\exp[-\pi(m_q^2+p_\perp^2)/\kappa]$. 
It should be noted, however, that this rate is only derived rigorously for constant, uniform abelian fields; its application to non-abelian QCD flux tubes is an analogy. 
Additionally, the Schwinger mechanism operates at the level of string break \emph{vertices}, not hadrons. 
Each vertex produces a $q'\bar{q}'$ pair with transverse momentum $\pm p_{\perp,k}$, and the hadron formed between adjacent vertices $k$ and $k+1$ carries the vector difference $p_{\perp,\text{hadron}} = p_{\perp,k} - p_{\perp,k+1}$. 

\textsc{Pythia} implements this picture by sampling the transverse momentum components $p_{x,k}$ and $p_{y,k}$ of each string breakup vertex independently from Gaussian distributions with width $\sigma$, treated as a tuneable parameter. 
The corresponding joint distribution for the transverse momentum at a given vertex is
\begin{equation}
  p(p_{x,k},p_{y,k})
  \;\propto\;
  \exp\!\left(-\frac{p_{x,k}^2+p_{y,k}^2}{2\sigma^2}\right),
  \label{eq:pythia_pT_components}
\end{equation}
Equivalently, this prescription induces a rotationally invariant distribution for the transverse momentum magnitude
$p_{\perp,k}=\sqrt{p_{x,k}^2+p_{y,k}^2}$,
\begin{equation}
  p(p_{\perp,k})
  \;\propto\;
  p_{\perp,k}\,
  \exp\!\left(-\frac{p_{\perp,k}^2}{2\sigma^2}\right),
  \label{eq:pythia_pT_radial}
\end{equation}
where the prefactor arises from the Jacobian in two transverse dimensions. 
The longitudinal momentum fraction $z$ is then drawn from \cref{eq:lund_ff_conditional}, with the resulting hadron-level transverse mass $m_\perp^2$ held fixed. 
This vertex-based construction implicitly induces correlations between the transverse momenta of neighboring hadrons through their shared string breakup vertices.

In this work, we instead adopt a hadron-level description in which each emission is specified directly by the kinematics of the produced hadron. 
We promote the conditional distribution \cref{eq:lund_ff_conditional} to a joint distribution over $(z,m_\perp^2)$,
\begin{equation}
  f(z, m_\perp^2)\,\mathrm{d}z\,\mathrm{d}m_\perp^2
  \;\propto\;
  \frac{(1-z)^a}{z}\,
  \exp\!\left( -b\,\frac{m_\perp^2}{z} \right)\mathrm{d}z\,\mathrm{d}m_\perp^2,
  \label{eq:lund_ff}
\end{equation}
which is understood to be supported on the physical region $0<z<1$ and $0<m_\perp^2<zM^2$. 
In this formulation, the marginal distribution of $m_\perp^2$ is determined by integrating \cref{eq:lund_ff} over $z$, rather than being imposed independently. 
Importantly, this choice leaves the conditional distribution $f(z|m_\perp^2)$ unchanged. 

The transverse dynamics stemming from the joint distribution in \cref{eq:lund_ff} represents a theoretically motivated departure from the Schwinger dynamics typically adopted in the Lund model. 
In \textsc{Pythia}, transverse momenta are sampled at string break vertices, so that adjacent hadrons share a vertex and carry correlated $p_T$. 
This convolution produces an approximately Gaussian hadron-level transverse momentum distribution and, crucially for practical implementations, the transverse recoil of the remnant string is set by the most recent break vertex momentum, remaining bounded and multiplicity-independent.
In the hadron-level kernel adopted here, successive emissions are statistically independent, yielding a non-Gaussian marginal distribution for $m_\perp^2$ with heavier tails governed by $\exp(-b\,m_\perp^2)$. 
An additional consequence is that the remnant transverse recoil $\vec{p}_{T,\rm rem} = -\sum_i \vec{p}_{T,i}$ executes a two-dimensional random walk whose variance grows linearly with multiplicity. 
The implications of this accumulated recoil for the choice of dynamical variable are discussed in \cref{sec:string_mass_update}. 

The hadron-level formulation is natural for the stochastic framework developed in this paper, in which the remaining string mass $M$ is the fundamental dynamical variable and each hadron emission corresponds to a single Markov step. 
As we will see in \cref{subsec:power_counting}, the full joint distribution admits a factorization in appropriately scaled coordinates under which the longitudinal and transverse momentum variables become statistically independent in the chiral limit. 
This factorization property is the foundation for the systematic power counting developed in the remainder of this work.

\subsection{String mass evolution}
\label{sec:string_mass_update}

For the purposes of constructing an effective description, it is sufficient to characterize the state of the hadronizing string by its \emph{string mass} $M$. 
Given a string of mass $M$, the production of a hadron with parameters $(z,m_\perp^2)$ reduces the string mass to
\begin{equation}\label{eq:string_mass_update}
  M'^2
  \;\equiv\; \Pi'^+\Pi'^-
  \;=\;
  \left( M^2 - \frac{m_\perp^2}{z} \right)(1 - z),
\end{equation}
where $\Pi'^\pm$ are the light-cone momenta of the remaining string after the emission. 
For a string initially at rest, $M^2$ coincides with the Lorentz-invariant mass squared.  After the first emission the remaining system carries transverse recoil $\vec{p}_T$, and the two quantities differ by $M_{\rm inv}'^2 = M'^2 - p_T^2$. 
Since string breaks are spacelike-separated on the worldsheet, transverse momentum generated at one break vertex cannot influence subsequent breaks, and the kinematic bound at each step, $m_\perp^2 \leq z M^2$, depends only on $\Pi^+\Pi^-$.
The string mass is therefore the natural one-dimensional Markov state variable for iterative fragmentation. 
The $p_T^2$ correction enters only at the termination boundary, where the physical criterion for cluster decay depends on $M_{\rm inv}$ rather than $M$ (see \cref{sec:endpoint_constraints}). 
A complete treatment would promote the remnant transverse recoil to a dynamical variable and condition on both $M$ and $p_{T,\rm rem}$ simultaneously, generating emergent forces in both directions through the Doob $h$-transform. 
For simplicity, the present work instead develops the EFT framework in the one-dimensional projection onto $M^2 = \Pi^+\Pi^-$, in which all transport coefficients and conditioned dynamics are exact. 
This projection is equivalent to the assumption that the accumulated remnant recoil satisfies $p_{T,\rm rem}^2 \ll M^2$ throughout the cascade---a condition automatically realized in the vertex-based transverse sector of \textsc{Pythia}, and expected to hold dynamically in the full two-dimensional conditioned process. 
The two-dimensional extension is deferred to future work.

For simplicity, we assume fragmentation proceeds from a single endpoint (the $q$ end) of the string. 
Fragmenting iteratively from both endpoints, as implemented in practical Monte Carlo generators, introduces no conceptual changes to the stochastic framework developed here; the bare kernel and mass update retain the same functional form, but the bookkeeping becomes more involved due to the need to track two independent fragmentation fronts and their eventual merger in the joining region. 
The single-ended treatment captures all essential features of the EFT tower construction while avoiding notational overhead. 

Together with the Lund kernel in \cref{eq:lund_ff}, this defines a discrete Markov chain whose transition kernel is given by
\begin{equation}
  \mathcal{K}_{\text{bare}}(M \to M')
  \;=\;
  \int \mathrm{d}z\,\mathrm{d}m_\perp^2\;
  f(z \mid m_\perp^2)\,p(m_\perp^2)\,
  \delta\!\left(M'^2 - \left( M^2 - \frac{m_\perp^2}{z} \right)(1 - z)\right),
  \label{eq:bare_mass_kernel}
\end{equation}
where $p(m_\perp^2)$ denotes the induced transverse-mass distribution at each string break. 

The continuum limit of this discrete evolution follows the general construction of \cref{subsec:continuum_limit}. 
When the string mass $M$ is large compared to typical hadronic scales, $M \gg \langle m_\perp \rangle \sim 0.5~\text{GeV}$, each fragmentation step removes only a small fraction of the total mass. 
Provided the transport coefficients satisfy $|\mu|, \sigma \ll M$, the Kramers--Moyal expansion of the discrete kernel truncates to a drift-diffusion generator (\cref{eq:bare_generator})
\begin{equation}
  \mathcal{L}
  \;=\;
  \mu(M)\,\partial_M
  \;+\;
  \frac{1}{2}\,\sigma^2(M)\,\partial_M^2,
  \label{eq:hadronization_generator}
\end{equation}
with transport coefficients determined by cumulants of the bare kernel:
\begin{equation}
  \mu(M) = \big\langle \Delta M \big\rangle_{\text{bare}}(M),
  \qquad
  \sigma^2(M) = \big\langle (\Delta M)^2 \big\rangle_{\text{bare}}(M) - \mu^2(M),
  \quad
  \Delta M \equiv M' - M.
  \label{eq:transport_coefficients}
\end{equation}
All microscopic details of the hadronization model enter the effective dynamics solely through these coefficients. 
The expansion breaks down as $M$ approaches hadronic scales (see \cref{subsec:boundary_EFT} for the treatment of this regime).

\subsection{Termination and the breakdown of the string picture}
\label{sec:endpoint_constraints}

The stochastic fragmentation dynamics defined by \cref{eq:hadronization_generator} cannot proceed indefinitely.
Two mass scales govern the termination of the iterative string-breaking description. 
The first, $M_\star$, is the minimum viable string mass for successful cluster termination.
A naive dimensional estimate would set $M_\star \sim \sqrt{\kappa} \approx 0.45$ GeV from the fundamental string scale, however, in practice the expected kinematics of a final two-body cluster decay can be used. 
The remnant cluster must balance the transverse momentum $\langle p_T \rangle$ carried by the final emitted hadron while producing two on-shell pions. 
This yields the threshold
\begin{equation}
  M_\star = \sqrt{(2m_\pi)^2 + \langle p_T \rangle^2},
  \label{eq:M_star_threshold}
\end{equation}
where the average transverse momentum is determined by the marginalized Lund kernel, $\langle p_T \rangle = \sqrt{1/(b(a+2))}$ (see appendix~\ref{app:moments}).
For Monash \cite{Skands:2014pea} parameters\footnote{The Monash parameters were tuned to \textsc{Pythia}'s sequential sampling procedure described by \cref{eq:lund_ff_conditional,eq:pythia_pT_radial}, not to the joint theoretical distribution in \cref{eq:lund_ff}. 
A dedicated tune using the full joint kernel would yield different optimal values for $a$ and $b$. 
We use Monash values throughout for concreteness and to facilitate comparison with existing implementations.} with $a = 0.68$, $b = 0.98$ GeV$^{-2}$, and $m_\pi = 0.14$ GeV, this gives $M_\star \approx 0.68$ GeV. 
The second scale, $M_{\rm cut}$, is a cutoff mass above $M_\star$ where the continuum approximation breaks down. 
Here the discrete hadronic spectrum becomes relevant and phase-space constraints tighten. 
In \textsc{Pythia}~\cite{Sjostrand:2006za,Sjostrand:2014zea}, fragmentation is terminated at $M_{\rm cut} \sim 1$--$2~\text{GeV}$, at which point an endpoint prescription (\textit{e.g.}~\texttt{finalTwo}) collapses the remnant into two on-shell hadrons.

From the perspective of the stochastic framework developed in \cref{sec:stoch_dynamics}, the termination requirement defines a concrete realization of the global constraint $\mathcal{C}$. 
The fragmentation trajectory must enter the \emph{success band}
\begin{equation}
  \mathcal{S} \equiv [M_\star, M_{\rm cut}]
  \label{eq:success_band}
\end{equation}
before crossing into the \emph{failure region}\footnote{The joint kernel adopted in this work enforces $m_\perp^2 \leq z M^2$ at each step, guaranteeing $M'^2 \geq 0$ and restricting the failure region to $\mathcal{F} = (0, M_\star)$. 
In \textsc{Pythia}'s sequential implementation the constraint $m_\perp^2/z \leq M^2$ is not imposed, so $M'^2$ can become negative and $\mathcal{F} = (-\infty, M_\star)$.}
\begin{equation}
  \mathcal{F} \equiv (0, M_\star).
  \label{eq:failure_region}
\end{equation}
Trajectories allowed by the bare kernel may be forbidden globally if they cannot terminate successfully, inducing non-Markovian correlations. 
Within the success band, the string description is replaced by an exclusive cluster decay. 

In practice, reaching $\mathcal{S}$ does not guarantee successful termination. 
As noted in \cref{sec:string_mass_update}, the physical criterion for cluster decay depends on the invariant mass $M_{\rm inv} = \sqrt{M^2 - p_T^2}$ rather than the string mass $M$, so that transverse recoil accumulated during fragmentation reduces the available phase space for the final decay. 
Additional constraints such as flavor or charge conservation may further reduce the cluster decay efficiency. 
These effects can be incorporated systematically by replacing the boundary condition $h(M) = 1$ for $M \in \mathcal{S}$ with an efficiency $h(M) = \eta(M) \leq 1$ that encodes the probability of successful cluster decay at a given string mass. 
This modifies the boundary data for the success probability but leaves the EFT structure unchanged. 
Throughout this work, we adopt unit efficiency $\eta = 1$ for simplicity, deferring a quantitative treatment of these corrections and systematic matching to an endpoint cluster model to future work. 
With the bare dynamics and termination constraints established, we now develop a systematic description of how these interact across different mass scales. 

\section{An effective tower for hadronization}\label{sec:effective_had_models}

The continuum limit of the bare fragmentation kernel, established in \cref{sec:hadronization}, reduces string mass evolution to a drift-diffusion process with transport coefficients $\mu(M)$ and $\sigma^2(M)$ determined by moments of the microscopic kernel. 
In this section we use this continuum limit to reveal an effective theory description of the bare fragmentation process, organizing the dynamics as a tower of effective theories valid across different ranges of string mass.

The construction proceeds in three steps. 
First, we establish a power-counting scheme based on the moment hierarchy of the Lund kernel, distinguishing longitudinal momentum sharing (controlled by the parameter $a$) from transverse phase space (suppressed by powers of $(bM^2)^{-1}$ at large $M$).
Second, we derive local effective theories governing typical fragmentation steps, identifying three kinematic regimes: an ultraviolet (UV) fixed point at asymptotically large $M$ where the dynamics are scale-invariant, an intermediate running regime where transverse effects induce controlled scale-dependence in the Wilson coefficients, and a boundary regime near $M_{\rm cut}$ where local and non-local effects become comparable. 
Third, we introduce explicit tail operators that encode rare, order-one jumps in the string mass---effects that lie outside the scope of the local diffusion approximation but are essential for computing global observables.

Importantly, the EFT tower constructed in this section describes the \emph{unconditioned} bare fragmentation dynamics. 
Global constraints such as exact energy-momentum conservation, which require the cascade to terminate successfully within the band $\mathcal{S} = [M_\star, M_{\rm cut}]$ rather than undershooting into the failure region $\mathcal{F}$, are not yet imposed. 
As discussed in \cref{sec:stoch_dynamics}, these constraints induce a renormalization of the effective generator through conditioning, a construction deferred to \cref{sec:h_global,sec:doob_h}.
The present section establishes the baseline dynamics against which the effects of global constraints can be systematically computed.

\subsection{Microscopic dynamics and power counting}\label{subsec:power_counting}

We begin by establishing the power-counting scheme that underlies the EFT expansion. 
As described in \cref{sec:hadronization}, each emitted hadron carries a fraction of light-cone momentum $z$ sampled from the bare fragmentation kernel
\begin{equation}\label{eq:lund_ff_recap}
  f(z, m_\perp^2) \propto z^{-1} (1-z)^a e^{-b m_\perp^2 / z}
\end{equation}
with $a,b > 0$, and kinematic constraints
\begin{equation}\label{eq:physical_mT_constraint}
  0 < z < 1, \quad m_\star^2 \leq m_\perp^2 \leq z M^2
\end{equation}
where $m_\star$ is the mass of lightest hadron available for production at a given fragmentation step. 

\subsubsection{Scaled coordinates and longitudinal-transverse coupling}

To expose the scaling structure of the kernel, we introduce the dimensionless variables\footnote{The variable $v = m_\perp^2/(zM^2)$ is motivated from the perspective of the string mass update in \cref{eq:string_mass_update}, where $m_\perp^2/z$ appears as the effective depletion of the longitudinal phase space.}
\begin{equation}
  u \equiv z, \quad v \equiv \frac{m_\perp^2}{z M^2}
\end{equation}
with the physical constraints in \cref{eq:physical_mT_constraint} mapping to
\begin{equation}\label{eq:uv_constraints}
  u \in (0,1), \quad v \in \left[\frac{m_\star^2}{u M^2}, 1\right].
\end{equation}
The $u$-dependent lower bound on $v$ introduces a mild coupling between longitudinal momentum sharing and transverse phase space. 
This coupling is parametrically small at large $M$, suppressed by $m_\star^2/M^2$, but grows toward the infrared boundary---characteristic of a relevant operator that mixes the UV and IR sectors in the effective theory expansion.

\subsubsection{The chiral limit and factorization}
In the limit $m_\star \to 0$, the constraint in \cref{eq:uv_constraints} relaxes to $v \in [0,1]$ and the coupling vanishes identically. 
The Lund kernel then exhibits exact \emph{longitudinal-transverse factorization} (see appendix~\ref{app:moments} for more details):
\begin{equation}\label{eq:factorization}
  f(z, m_\perp^2) \,\mathrm{d}z\, \mathrm{d}m_\perp^2
  \;=\;
  p(u) \, p(v \mid M) \, \mathrm{d}u\, \mathrm{d}v,
\end{equation}
where
\begin{equation}\label{eq:factorized_distributions}
  p(u) = (1+a)(1-u)^a, \quad \quad p(v \mid M) = \frac{bM^2 e^{-bM^2v}}{1-e^{-bM^2}}.
\end{equation}
For a given string mass $M$, the variables $u$ and $v$ become statistically independent: $p(u,v \,| \, M) = p(u) \, p(v \,|\, M)$. 
The longitudinal momentum fraction is governed purely by the left-right symmetry parameter $a$, while the scaled transverse phase space is governed by the dimensionless combination $\lambda \equiv bM^2$.

Physically, this factorization is exact in the chiral limit $m_q \to 0$ of QCD, 
where pions become massless Nambu-Goldstone bosons of spontaneous chiral symmetry breaking. 
The chiral limit removes explicit breaking terms proportional to $m_q$, ensuring $m_\star \equiv m_\pi = 0$ while preserving the full hadronic spectrum generated by dynamical symmetry breaking. 
The confinement scale $\Lambda_{\rm QCD}$ and string tension $\kappa$ arise from nonperturbative gluon dynamics rather than quark masses, and are therefore unaffected by the chiral limit. 

While not strictly required for the EFT construction, this factorization provides substantial analytic simplification. 
The statistical independence ensures that all mixed moments factorize exactly,
\begin{equation}
  \langle u^p v^q \rangle = \langle u^p \rangle \langle v^q \rangle,
\end{equation}
cleanly separating UV fragmentation physics (encoded in moments of $u$) from IR transverse phase space (encoded in moments of $v$). 
Retaining the finite-$m_\star$ coupling would introduce mixing between these sectors at the level of the moment expansion, obscuring the parametric hierarchy between longitudinal and transverse contributions. 
We therefore work in the chiral limit throughout this paper. 
The systematic treatment of finite-mass corrections as relevant deformations will be addressed in future work. 
Note that in the chiral limit the termination scale in eq.~\eqref{eq:M_star_threshold} simplifies to $M_\star = \langle p_T \rangle \approx 0.62$ GeV.

\subsubsection{Moment hierarchy and power counting}

The factorization in \cref{eq:factorization} establishes a manifest moment hierarchy. 
The longitudinal moments $\langle u^p \rangle$ are determined by the $\mathcal{O}(1)$ Lund parameter $a$, with explicit expressions given in appendix~\ref{app:moments}. 
The transverse moments $\langle v^q \rangle$ scale as $(bM^2)^{-q} \equiv \lambda^{-q}$ and vanish parametrically at large mass. 
At large $M$, longitudinal dynamics are $\mathcal{O}(1)$ while transverse corrections enter as controlled powers of $\lambda^{-1} = (bM^2)^{-1} \ll 1$.

From \cref{eq:string_mass_update}, the $n$-th ($M_{n} \to M_{n+1}$) string mass update also factorizes accordingly
\begin{equation}\label{eq:power_counting_exp}
  M_{n+1} = M_n \sqrt{1 - u_n}\sqrt{1 - v_n}.
\end{equation}
The mass increment then admits a systematic expansion in the random variables controlling longitudinal and transverse momentum
\begin{equation}\label{eq:bivariate_expansion}\begin{aligned}
  \frac{\Delta M}{M} \equiv \frac{M_{n+1} - M_{n}}{M_n} &= \left( 1 - \frac{1}{2} u_n - \frac{1}{8} u_n^2 - \cdots \right) \left( 1 - \frac{1}{2} v_n - \frac{1}{8} v_n^2 - \cdots \right) - 1 \\[0.1in]
  &=\sum_{p,q \geq 0} c_{pq} \, u_n^p \, v_n^q
\end{aligned}\end{equation}
with coefficients $c_{pq}$ fixed by the binomial series. 

This moment hierarchy provides the foundation for the local effective theories constructed below. 
Expectation values of smooth functionals of $\Delta M$ may be systematically approximated by truncating \cref{eq:bivariate_expansion} at fixed order in $(p,q)$, defining a local EFT organized around the leading moments of the longitudinal and transverse fragmentation variables.

\subsection{Exact drift and diffusion from the microscopic kernel}\label{subsec:local_eft}

\begin{figure}[t]
\centering
\includegraphics[width=1.0\textwidth]{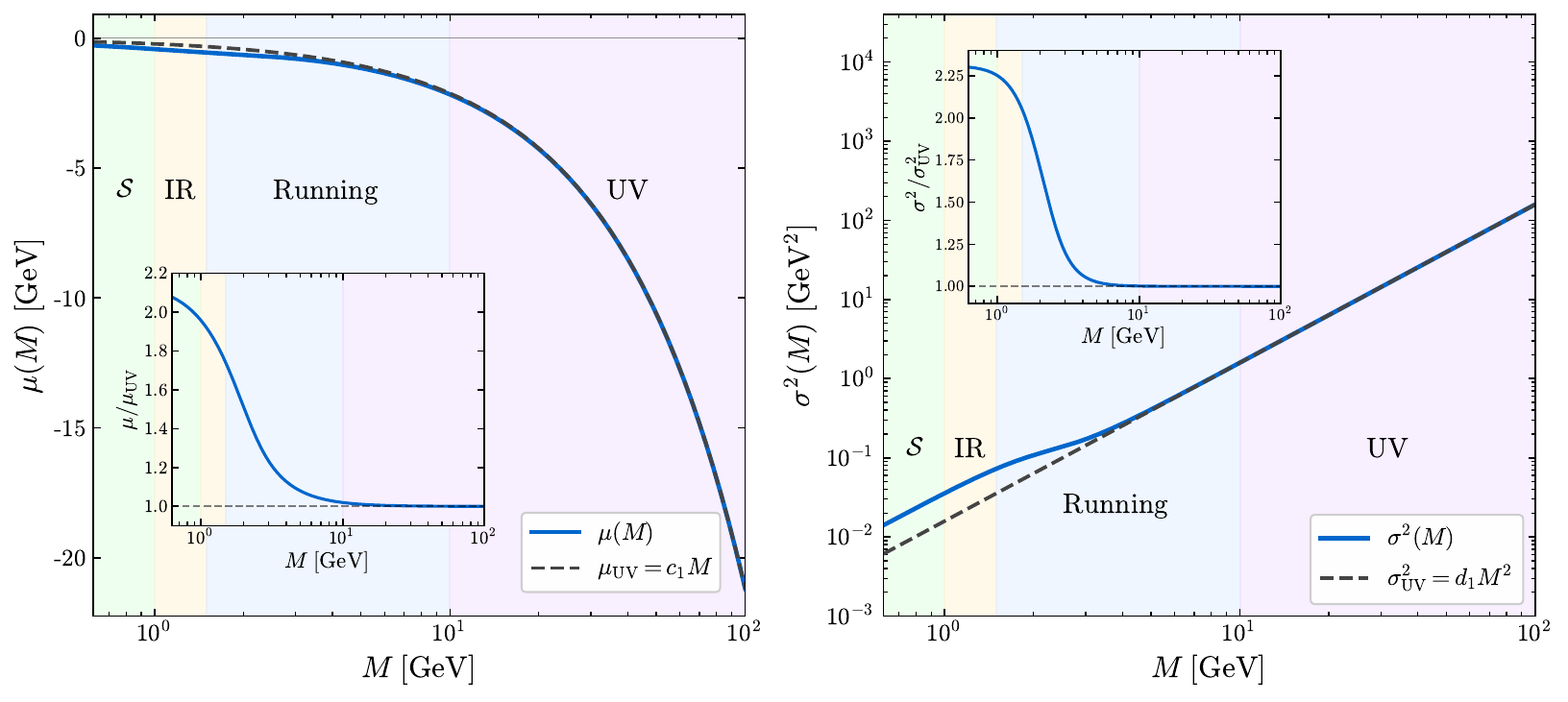}
\caption{Transport coefficients governing the continuum limit of string mass evolution. \textbf{Left:} Drift coefficient $\mu(M)$ (mean mass loss per step) and its UV fixed-point $\mu_{\rm UV} = c_1 M$ with $c_1 = -1/(2a+3) \approx -0.23$. \textbf{Right:} Diffusion coefficient $\sigma^2(M)$ (variance of mass fluctuations) and UV scaling $\sigma^2_{\rm UV} = d_1 M^2$. Insets show ratios approaching unity for $M \gtrsim 10$ GeV, confirming scale-invariant UV behavior. Shaded regions denote the EFT tower: termination band $\mathcal{S}$ ($M_\star < M < M_{\rm cut}$ with $M_\star = 0.62$ GeV, $M_{\rm cut} = 1.0$ GeV), IR boundary layer ($1.0$\,--\,$1.5$ GeV), running regime ($1.5$\,--\,$10$ GeV), and UV fixed point ($M > 10$ GeV). Monash tune: $a = 0.68$, $b = 0.98$ GeV$^{-2}$.}
\label{fig:drift_diffusion}
\end{figure}

Following the continuum construction of \cref{subsec:continuum_limit,sec:string_mass_update}, the drift and diffusion coefficients governing the local evolution are determined by the first two moments of the mass increment. 
Using the microscopic update $M_{n+1}=M\sqrt{(1-u)(1-v)}$, the drift and diffusion coefficients admit
exact representations
\begin{equation}
  \mu(M)
  =
  M\,
  \Big\langle \sqrt{(1-u)(1-v)} - 1 \Big\rangle,
  \label{eq:mu_exact}
\end{equation}
\begin{equation}
  \sigma^2(M)
  =
  M^2\,
  \Big\langle \big(\sqrt{(1-u)(1-v)} - 1\big)^2 \Big\rangle
  - \mu(M)^2,
  \label{eq:sigma_exact}
\end{equation}
where the expectation values are taken with respect to the exact joint distribution of $(u,v)$ induced by the Lund kernel, with $u$ and $v$ sampled independently at fixed $M$. 
\Cref{eq:mu_exact,eq:sigma_exact} are valid for all $M$ and contain no approximation beyond the assumption of independent fragmentation steps. 
General identities for moments of $(1-u)(1-v)$ are collected in appendix~\ref{app:moments}.

In practice, the exact expressions \cref{eq:mu_exact,eq:sigma_exact} are most transparently organized using the expansion of $\Delta M/M$ derived in \cref{eq:bivariate_expansion}. 
Expanding $\Delta M/M$ in moments of $u$ and $v$ and retaining the leading terms yields
\begin{equation}\label{eq:mu_moment}
  \frac{\mu(M)}{M}
  =
  -\frac12\!\left(\langle u\rangle + \langle v\rangle\right)
  + \frac14\langle uv\rangle
  - \frac18\!\left(\langle u^2\rangle + \langle v^2\rangle\right)
  + \cdots,
\end{equation}
\begin{equation}\label{eq:sigma_moment}
  \frac{\sigma^2(M)}{M^2}
  =
  \frac14\!\left(\langle u^2\rangle + \langle v^2\rangle
  + 2\langle uv\rangle\right)
  - \frac14\!\left(\langle u\rangle + \langle v\rangle\right)^2
  + \cdots.
\end{equation}
In appendix~\ref{app:moments} we show that 
\begin{equation}\label{eq:u_moments}
  \langle u^{p} \rangle = \mathbb{E}[u^p] = \dfrac{p!}{\prod_{k=2}^{p+1}(a+k)}
\end{equation}
\begin{equation}\label{eq:v_moments}
  \langle v^{q} \rangle = \mathbb{E}[v^q] = \dfrac{q!}{\lambda^q (1 - e^{-\lambda})}\left[1 - e^{-\lambda} \sum_{k=0}^{q} \frac{\lambda^k}{k!} \right]
\end{equation}
where, again, $\lambda \;\equiv\; b M^2$ controls the transverse-momentum phase space and governs the $M$-dependence of the $v$-moments.

The moment expressions in \cref{eq:mu_moment,eq:sigma_moment}, together with the exact formulas for $\langle u^p \rangle$ and $\langle v^q \rangle$, determine the transport coefficients $\mu(M)$ and $\sigma^2(M)$ across all mass scales. 
The parametric hierarchy $\langle v^q \rangle \sim \lambda^{-q}$ implies that these coefficients exhibit distinct behavior in three kinematic regimes: an ultraviolet fixed point at large $M$ where transverse effects are negligible, an intermediate running regime where finite-$\lambda$ corrections induce controlled scale dependence, and a boundary layer near $M_{\rm cut}$ where the continuum approximation begins to break down. 
We now construct local effective theories adapted to each regime; the resulting scale dependence of the transport coefficients is illustrated in \cref{fig:drift_diffusion}.

\subsubsection{UV theory -- a local fixed-point}
\Cref{eq:mu_moment,eq:sigma_moment} define a hierarchy of local effective theories corresponding to controlled truncations of the same generator \cref{eq:hadronization_generator}. 
In the ultraviolet regime $\lambda \gg 1$, moments of the transverse variable $v$ are parametrically suppressed and the local generator reduces to a scale-invariant fixed point governed entirely by longitudinal momentum sharing,
\begin{equation}
  \mu_{\rm UV}(M)=c_1\,M,
  \qquad
  \sigma^2_{\rm UV}(M)=d_1\,M^2.
\end{equation}
In the language of effective field theory, $c_1$ and $d_1$ are Wilson coefficients determined entirely by moments of the longitudinal fragmentation variable $u$. 
The explicit evaluation of these coefficients from the longitudinal Lund kernel, using the moment expansion in \cref{eq:mu_moment,eq:sigma_moment,eq:u_moments}, is straightforward and presented in \cref{app:moments}. 
We identify
\begin{equation}
  \begin{aligned}
  c_1 &\simeq \langle \sqrt{1-u} - 1 \rangle = - \frac{1}{2a + 3}
  \end{aligned}\end{equation}
\begin{equation}\begin{aligned}
  d_1 &\simeq \langle (\sqrt{1-u} - 1)^2 \rangle - \langle \sqrt{1-u} - 1 \rangle^2= \frac{a+1}{(a+2)(2a+3)^2}.
\end{aligned}\end{equation}
This UV local EFT captures ``free'' hadronization dynamics far from the termination region and uninfluenced by phase space or global constraints.

\vspace{0.5em}
\subsubsection{Intermediate running theory -- finite mass effects}

At finite but still large mass, moments of the transverse variable $v=m_\perp^2/(uM^2)$ generate controlled corrections to the ultraviolet fixed point. 
Retaining the same local operator basis as in the UV theory, but keeping the full $\lambda=bM^2$ dependence of the $v$-moments, defines an intermediate local EFT in which the drift and diffusion coefficients acquire mild, computable scale dependence.

Concretely, we write
\begin{equation}
  \mu_{\rm int}(M)=M\,C_1(\lambda),
  \qquad
  \sigma^2_{\rm int}(M)=M^2\,D_1(\lambda),
  \qquad
  \lambda\equiv bM^2,
\end{equation}
where $C_1(\lambda)$ and $D_1(\lambda)$ are running Wilson coefficients whose scale dependence is fixed uniquely by the cumulant expansion of the microscopic fragmentation kernel. 
At fixed truncation order in the moment expansion introduced in \cref{subsec:local_eft}, the functions $C_1(\lambda)$ and $D_1(\lambda)$ resum the full $\lambda=bM^2$ dependence of the transverse phase space. 
Using the exact moments of $v$ from \cref{eq:v_moments}, this yields the closed-form expressions
\begin{equation}
  C_1(\lambda) = A_1 B_1(\lambda) - 1
\end{equation}
\begin{equation}
  D_1(\lambda) = A_2 B_2(\lambda) - (A_1 B_1(\lambda))^2
\end{equation}
with
\begin{equation}
  A_1 = \frac{a+1}{a + 3/2}, \qquad B_1(\lambda) = \frac{1}{1-e^{-\lambda}} \left[ 1-\frac{e^{-\lambda} \sqrt{\pi}}{2\sqrt{\lambda}} \text{erfi}(\sqrt{\lambda}) \right]
\end{equation}
\begin{equation}
  A_2 = \frac{a+1}{a+2}, \qquad B_2(\lambda) = 1-\frac{1}{\lambda} + \frac{e^{-\lambda}}{1-e^{-\lambda}}.
\end{equation}
Again, the derivation of these expressions is reserved for \cref{app:moments}. 
In the limit $\lambda\to\infty$, all moments of $v$ vanish and the intermediate theory reduces smoothly to the ultraviolet local EFT,
\begin{equation}
  \lim_{\lambda\to\infty} C_1(\lambda)=c_1,
  \qquad
  \lim_{\lambda\to\infty} D_1(\lambda)=d_1.
\end{equation}
The intermediate local EFT therefore provides a controlled, analytic interpolation between the UV fixed point and the onset of boundary effects, while remaining strictly local in the string mass.

\subsubsection{IR/Boundary theory -- local EFT near \texorpdfstring{$M_{\text{cut}}$}{M\_cut}}
\label{subsec:boundary_EFT}

Near the termination scale $M_{\rm cut}$, the mass evolution departs from the ultraviolet and intermediate regimes in that both local fluctuations in $M$ and rare, order--one fragmentation steps become relevant. 
In this subsection we introduce the \emph{local} component of the boundary EFT, which describes the continuation dynamics in the vicinity of $M_{\rm cut}$. 
Genuinely non--local effects associated with one--step boundary crossings are deferred to \cref{subsec:nonlocal_operators}.

The local boundary EFT is defined by expanding the exact drift and diffusion coefficients about the cutoff scale,
\begin{equation}
  \delta M \equiv M - M_{\rm cut}.
\end{equation}
Expanding
\begin{align}
  \mu_{\rm IR}(M)
  &= c_0^{\rm IR}
   + c_1^{\rm IR}\,\delta M
   + \tfrac12 c_2^{\rm IR}\,\delta M^2 + \cdots,
  \\
  \sigma^2_{\rm IR}(M)
  &= d_0^{\rm IR}
   + d_1^{\rm IR}\,\delta M
   + \tfrac12 d_2^{\rm IR}\,\delta M^2 + \cdots,
\end{align}
with Wilson coefficients $\{c_i^{\rm IR}, d_i^{\rm IR}\}$ fixed by derivatives of the
exact microscopic drift and diffusion evaluated at $M=M_{\rm cut}$
\begin{equation}
  c_n^{\rm IR}
  =
  \left.\frac{d^n\mu}{dM^n}\right|_{M_{\rm cut}},
  \qquad
  d_n^{\rm IR}
  =
  \left.\frac{d^n\sigma^2}{dM^n}\right|_{M_{\rm cut}}.
  \label{eq:boundary_Wilson_def}
\end{equation}
These expressions will be used below as matching data for global observables, but their physical role only becomes apparent once non--local effects are included. 

At leading order one finds
\begin{align}
  c_0^{\rm IR}
  &= M_{\rm cut}\big[A_1B_1(\lambda_c)-1\big],
  \\
  d_0^{\rm IR}
  &= M_{\rm cut}^2\big[A_2B_2(\lambda_c)-(A_1B_1(\lambda_c))^2\big],
\end{align}
with $\lambda_c=bM_{\rm cut}^2$.
The first derivative coefficients are
\begin{align}
  c_1^{\rm IR}
  &=
  \big[A_1B_1(\lambda_c)-1\big]
  +2bM_{\rm cut}^2A_1B_1'(\lambda_c),
  \\
  d_1^{\rm IR}
  &=
  2M_{\rm cut}D_1(\lambda_c)
  +2bM_{\rm cut}^3
  \Big[A_2B_2'(\lambda_c)
  -2A_1^2B_1(\lambda_c)B_1'(\lambda_c)\Big],
\end{align}
where primes denote derivatives with respect to $\lambda$. 
Higher--order coefficients follow analogously.

While the local boundary EFT provides a controlled description of the continuation dynamics near $M_{\rm cut}$, it is not, by itself, complete. 
A full effective description requires supplementing it with non-local structures that encode rare, order-one fragmentation events.

\subsection{Non-local operators and rare fragmentation events}\label{subsec:nonlocal_operators}

The local diffusion EFT constructed in the previous subsection governs the evolution of \emph{local observables}, whose one-step expectation values are controlled by infinitesimal changes in the string mass and admit a controlled Taylor expansion in $\Delta M$. 
Within its regime of validity, this description captures the cumulative effect of many typical fragmentation steps and provides a complete characterization of local transport through the drift and diffusion coefficients $\mu(M)$ and $\sigma^2(M)$.

There exist, however, physically important observables that lie outside the scope of the local EFT. 
A canonical example is the probability for the fragmentation cascade to cross a specified mass threshold in a single step, which plays a central role in the construction of conditioned dynamics in \cref{sec:h_global}. 
The local diffusion generator cannot capture such observables because it describes a \emph{continuous} stochastic process: reaching $M_{\rm cut}$ from large $M$ via diffusion requires passing through all intermediate mass values, accumulated over many small steps. 
The microscopic Lund dynamics, however, allows \emph{jumps}---a single fragmentation can carry the string mass directly from $M = 5$~GeV to $M' < M_{\rm cut}$, bypassing intermediate values entirely. 
No finite truncation of the Kramers--Moyal expansion can reproduce such discontinuous transitions.

Near the termination boundary, both processes contribute at order one. 
A trajectory may diffuse into $\mathcal{S}$ through many small steps, or jump there directly in a single fragmentation event. 
To account for both within the EFT framework, the local diffusion generator must be supplemented by explicit non-local structures that encode the probability for rare, order-one jumps. We now introduce these in the form of \emph{tail operators}, which quantify direct transitions across macroscopic regions in string mass.

\subsubsection{Tail operators}

\begin{figure}[t]
\centering
\includegraphics[width=1.0\textwidth]{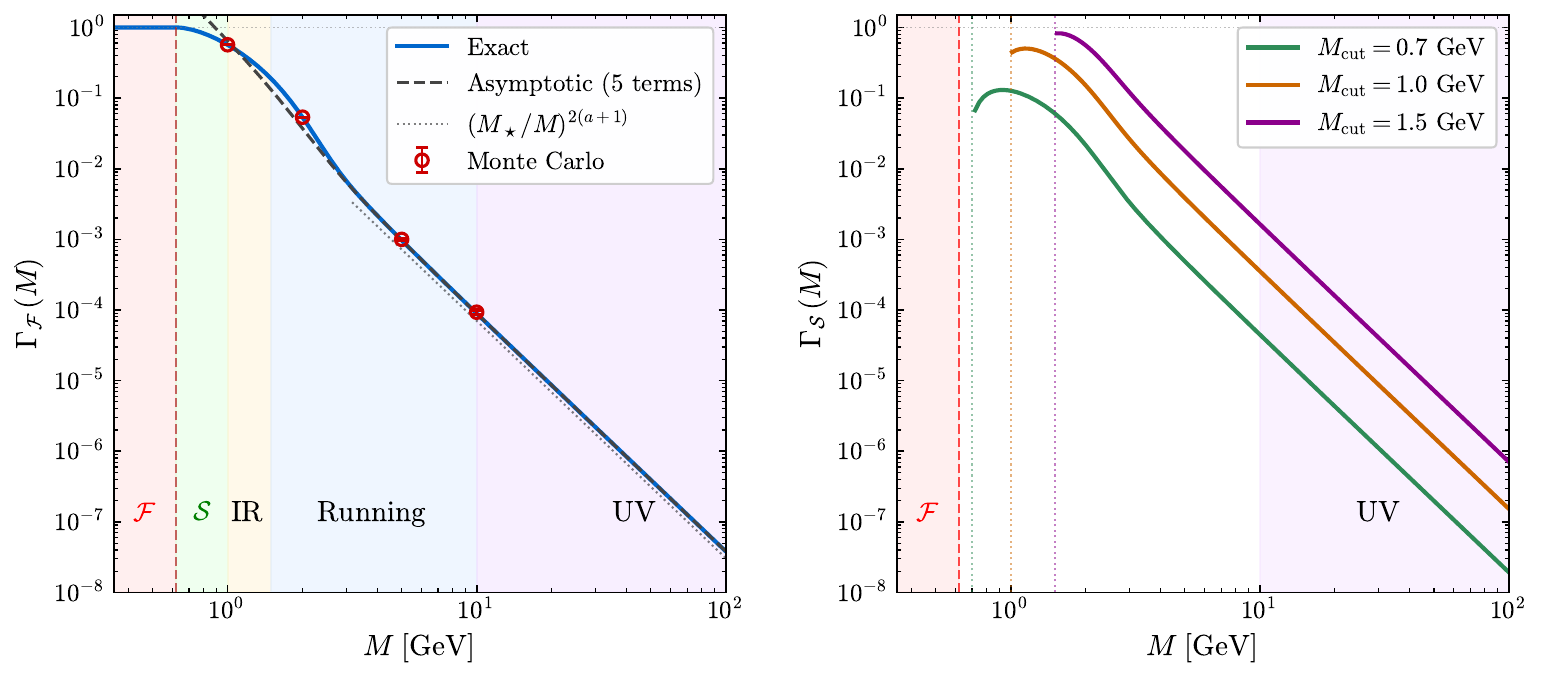}
\caption{Non-local tail operators encoding single-step transition probabilities. \textbf{Left:} Failure operator $\Gamma_{\mathcal{F}}(M)$, the probability of undershooting into $\mathcal{F} = (0, M_\star)$ with $M_\star = 0.62$ GeV. \textbf{Right:} Termination operator $\Gamma_{\mathcal{S}}(M)$, the probability of landing in $\mathcal{S} = [M_\star, M_{\rm cut}]$, shown for $M_{\rm cut} = 0.7, 1.0, 1.5$ GeV. Both scale as $(M_{\rm scale}/M)^{2(a+1)} \approx (M_{\rm scale}/M)^{3.4}$ in the UV and are power-suppressed at large $M$, but become $\mathcal{O}(1)$ near $M_{\rm cut}$ where they enter the EFT at leading order.}
\label{fig:tail_operators}
\end{figure}

We characterize non-local effects by defining tail operators that encode the probability for a single fragmentation step to cross specified mass thresholds. 
In particular, we define
\begin{align}
  \Gamma_{\mathcal{F}}(M)
  &\equiv
  \mathbb{P}\!\left(M_{n+1}<M_\star \mid M_n=M\right),
  \label{eq:GammaF_def}
  \\
  \Gamma_{\mathcal{S}}(M)
  &\equiv
  \mathbb{P}\!\left(M_\star \le M_{n+1}\le M_{\rm cut} \mid M_n=M\right),
  \label{eq:GammaR_def}
\end{align}
which encode the probabilities for a single fragmentation step to cross directly into the failure and termination regions, respectively.

Using the microscopic update $M_{n+1}=M\sqrt{(1-u)(1-v)}$,
the conditions in \cref{eq:GammaF_def,eq:GammaR_def} translate into
\begin{equation}
  (1-u)(1-v)
  \le
  \frac{M_\star^2}{M^2},
  \qquad
  \frac{M_\star^2}{M^2}
  \le
  (1-u)(1-v)
  \le
  \frac{M_{\rm cut}^2}{M^2}.
\end{equation}
The tail operators therefore admit exact integral representations
\begin{align}
  \Gamma_{\mathcal{F}}(M)
  &=
  \int_0^1 du \int_0^1 dv\;
  p(u)\,p(v\mid M)\;
  \Theta\!\left(
    \frac{M_\star^2}{M^2}
    -
    (1-u)(1-v)
  \right),
  \label{eq:GammaF_exact}
  \\
  \Gamma_{\mathcal{S}}(M)
  &=
  \int_0^1 du \int_0^1 dv\;
  p(u)\,p(v\mid M)\;
  \Theta\!\left(
    (1-u)(1-v)
    -
    \frac{M_\star^2}{M^2}
  \right)
  \Theta\!\left(
    \frac{M_{\rm cut}^2}{M^2}
    -
    (1-u)(1-v)
  \right).
  \label{eq:GammaR_exact}
\end{align}
Unlike the local drift and diffusion coefficients, these quantities probe the endpoint structure of the fragmentation kernel and are intrinsically non-local in $M$. 
For the Lund kernel, the integrals \cref{eq:GammaF_exact,eq:GammaR_exact} may be evaluated analytically (see appendix~\ref{app:tail_operators} for more details).
Introducing $\epsilon \equiv M_{\rm scale}^2/M^2$ and defining
\begin{equation}\begin{aligned}
  \mathcal{G}(\epsilon; M) =
  \frac{\lambda}{e^{\lambda}-1}
  \Big[\epsilon^{a+1} E_{a+1}(-\lambda) - \epsilon \, E_{a+1}(-\lambda\epsilon)\Big]
  \;+\;
  \frac{e^{\lambda\epsilon}-1}{e^{\lambda}-1},
  \label{eq:Gamma_closed}
\end{aligned}\end{equation}
where $E_\nu(z) = \int_1^\infty t^{-\nu} e^{-zt}\, dt$ is the generalized exponential integral. 
The tail operators are then given by
\begin{equation}
  \Gamma_{\mathcal{F}}(M)=\mathcal{G}(\epsilon_\star; M),
  \qquad
  \Gamma_{\mathcal{S}}(M)=\mathcal{G}(\epsilon_{\rm cut}; M)-\mathcal{G}(\epsilon_\star; M),
  \qquad
  \epsilon_\star=\frac{M_\star^2}{M^2},
  \quad
  \epsilon_{\rm cut}=\frac{M_{\rm cut}^2}{M^2}.
\end{equation}

At large string masses where $M \gg M_{\rm scale}$, the tail probabilities are strongly suppressed, scaling as
\begin{equation}
  \Gamma_{\mathcal{F},\mathcal{S}}(M) \sim \left(\frac{M_{\rm scale}}{M}\right)^{2(a+1)},
  \label{eq:Gamma_scaling}
\end{equation}
with subleading corrections organized as a series in the transverse moments $\langle v^n \rangle_\lambda \sim (bM^2)^{-n}$. 
The full asymptotic expansion and its derivation are presented in appendix~\ref{app:tail_operators}. 
In practice, we employ the exact integral representations \cref{eq:GammaF_exact,eq:GammaR_exact} when computing non-local observables; the asymptotic expansion is used primarily to establish power counting and demonstrate overlap with the local EFT. 
The behavior of these operators across the full range of string mass is shown in \cref{fig:tail_operators}.

The tail operators $\Gamma_{\mathcal{F}}(M)$ and $\Gamma_{\mathcal{S}}(M)$ occupy a distinct role within the EFT tower. 
Unlike the local Wilson coefficients $\mu(M)$ and $\sigma^2(M)$, which govern transport through many typical fragmentation steps, the tail operators encode rare single-step events that induce order-one changes in $M$. 
Instead, the tail operators enter the effective description only when considering \emph{non-local observables} that are sensitive to global properties of fragmentation trajectories. 
At the level of effective operators, the hadronization dynamics may be decomposed as
\begin{equation}
  \mathcal{L}_{\rm eff}
  =
  \mathcal{L}_{\rm loc}
  +
  \mathcal{L}_{\rm nonloc},
  \label{eq:Leff_split}
\end{equation}
where $\mathcal{L}_{\rm loc}$ is the local diffusion generator constructed in the preceding subsections. 
The non-local contribution acts as a jump operator into absorbing boundary sets,
\begin{equation}
  \mathcal{L}_{\rm nonloc}\,\mathcal O(M)
  =
  \Gamma_{\mathcal{F}}(M)\big[\mathcal O(M_\star)-\mathcal O(M)\big]
  +
  \Gamma_{\mathcal{S}}(M)\big[\mathcal O(M_{\rm cut})-\mathcal O(M)\big],
  \label{eq:L_nonloc_def}
\end{equation}
where $\Gamma_{\mathcal{F}}(M)$ and $\Gamma_{\mathcal{S}}(M)$ are the exact tail probabilities for direct one-step transitions into the failure and termination regions, respectively.

\subsubsection{The \texorpdfstring{$\delta$}{delta}-split and scheme dependence}
\label{subsubsec:tail_local_complementarity}

At large string mass, the local diffusion generator and tail operators describe complementary physics: local transport dominates typical fragmentation steps, while tail operators encode rare large jumps that are parametrically suppressed. 
Near the boundary $M \sim M_{\rm cut}$, however, this separation breaks down. 
Both descriptions become order-one, and since both derive from the same microscopic Lund kernel, their domains of validity overlap. 
A naive combination $\mathcal{L}_{\rm loc} + \mathcal{L}_{\rm nonloc}$ would double-count fragmentation events that contribute both to the moments defining $\mu(M)$ and $\sigma^2(M)$ and to the exact tail integrals.

The resolution is to partition phase space explicitly. 
Define the logarithmic step $\Delta x \equiv \ln(M'/M)$ and introduce a splitting scale $\delta > 0$. 
Steps with $|\Delta x| < \delta$ are treated via the Kramers--Moyal expansion; steps with $|\Delta x| > \delta$ are integrated exactly. 
This yields
\begin{equation}
  \mathcal{L}_{\rm eff} = \mathcal{L}_{\rm loc}(\delta) + \mathcal{L}_{\rm tail}(\delta),
  \label{eq:Leff_delta_split}
\end{equation}
where the transport coefficients $\mu(M,\delta)$ and $\sigma^2(M,\delta)$ are computed from restricted integrals over $|\Delta x| < \delta$, ensuring each fragmentation event contributes to exactly one sector.

The splitting scale $\delta$ is not physical but an organizational choice analogous to the renormalization scheme in quantum field theory. 
Physical observables are independent of $\delta$ because varying it merely redistributes the same underlying physics between sectors: increasing $\delta$ transfers probability flux from the tail to the local sector, with corresponding changes in $\mu$ and $\sigma^2$ that exactly compensate. 
Formally,
\begin{equation}
  \frac{\partial \mathcal{L}_{\rm eff}}{\partial \delta} = \mathcal{O}(\delta^3),
\end{equation}
with residual dependence entering only at the order of Kramers--Moyal truncation errors.

In practice, $\delta$ should be comparable to the typical step size $|\mu(M)|/M$ set by the drift coefficient, ensuring that ordinary diffusive transport remains in the local sector while genuine large jumps are captured by the tail operators. 
From the UV fixed point, $\mu_{\rm UV} = c_1 M$ with $c_1 = -1/(2a+3) \approx -0.23$ for Monash parameters, yielding $|\mu|/M \approx 0.4$. 
A choice of $\delta \approx 0.3$ is therefore appropriate, placing the partition at the boundary between typical diffusive steps and rare order-one jumps.

The effective description developed in this section---the local EFT tower together with the non-local tail operators---provides a complete characterization of the bare fragmentation dynamics. 
What it does not yet incorporate are global constraints on fragmentation histories. 
In the next section we construct global observables such as the success probability $h(M)$, which quantifies whether a cascade initiated at mass $M$ can terminate successfully. 
These observables induce a renormalization of the effective dynamics through conditioning, revealing how global constraints modify the coarse-grained evolution encoded in the EFT tower.

\section{Global observables}
\label{sec:h_global}

The local and non-local EFTs constructed in \cref{sec:effective_had_models} describe the \emph{bare} fragmentation dynamics: the string mass $M$ evolves as a Markov chain whose transition kernel is fixed by the Lund splitting law. 
As discussed in \cref{sec:stoch_dynamics}, in many physical applications the relevant ensemble is not the bare process itself but a \emph{conditioned} one, in which only fragmentation histories satisfying a global constraint are retained.

\subsection{The success probability}

In practice, successful termination requires that the fragmentation trajectory enter the \emph{termination band} $\mathcal{S} = [M_\star, M_{\rm cut}]$ before crossing into the \emph{failure region} $\mathcal{F} = (0, M_\star)$.
This defines a concrete realization of the global constraint $\mathcal{C}$ introduced in \cref{sec:stoch_dynamics,sec:hadronization}, with the termination requirement serving as a proxy for the ability to complete hadronization with exact energy-momentum and quantum-number conservation via a low-energy cluster decay.

The success probability
\begin{equation}
  h(M)
  =
  \mathbb{P}_{\rm bare}\!\left(
    \mathcal{C} \,\big|\, M_0=M
  \right)
  =
  \mathbb{P}_{\rm bare}\!\left(
    \text{reach } M\in\mathcal{S}
    \text{ before } M\in\mathcal{F}
    \,\big|\, M_0=M
  \right),
  \label{eq:h_def_h_sec}
\end{equation}
is a global first--passage observable of the bare fragmentation process.

Unlike the local observables considered in \cref{sec:effective_had_models}, the function $h(M)$ depends on the full future fragmentation history. 
It probes rare, order--one fragmentation events and the global structure of fragmentation trajectories, and therefore provides a sharp diagnostic of how the local and non-local components of the EFT combine in enforcing global termination constraints.

As discussed in \cref{sec:stoch_dynamics}, $h(M)$ also plays a central role in the Doob $h$-transform, where each transition $M \to M'$ is reweighted by the factor $h(M')/h(M)$, biasing the evolution toward successful configurations. 
Computing $h(M)$ therefore serves two purposes, providing both the matching observable for the EFT tower and the renormalization function needed to construct the conditioned dynamics in \cref{sec:doob_h}.

\subsection{Exact backward equation for \texorpdfstring{$h$}{h}} \label{subsec:exact_h}

Since $h(M)$ is the probability of eventual success starting from mass $M$, and the bare kernel $\mathcal{K}_{\rm bare}(M \to M')$ governs each fragmentation step, the law of total probability gives
\begin{equation}
  h(M) = \int_0^\infty \mathrm{d}M'\;
  \mathcal{K}_{\rm bare}(M \to M')\,h(M'),
  \label{eq:h_one_step}
\end{equation}
with absorbing boundary conditions
\begin{equation}\label{eq:h_bc}
  h(M)=1 \quad (M\in\mathcal{S}),
  \qquad
  h(M)=0 \quad (M\in\mathcal{F}),
\end{equation}
reflecting successful termination and irreversible failure, respectively. 
The condition $h(M) = 1$ on $\mathcal{S}$ assumes that the accumulated remnant recoil satisfies $p_{T,\rm rem}^2 \ll M_{\rm cut}^2$ at termination, so that $M_{\rm inv} \approx M$. 
A complete treatment would promote $p_{T,\rm rem}$ to a dynamical variable and condition on both $M$ and $p_{T,\rm rem}$ simultaneously, as discussed in \cref{sec:lund_kernel,sec:endpoint_constraints}.

A fragmentation step starting at $M > M_{\rm cut}$ either exits to the termination band $\mathcal{S}$, undershoots into the failure region $\mathcal{F}$, or remains in the continuation region $M' > M_{\rm cut}$. 
Applying the boundary conditions \cref{eq:h_bc} to the exit channels gives
\begin{equation}
  h(M) = \Gamma_{\mathcal{S}}(M)
         + \int_{M_{\rm cut}}^\infty \mathcal{K}_{\rm bare}(M \to M')\,h(M')\,\mathrm{d}M',
  \label{eq:h_integral_eq}
\end{equation}
where $\Gamma_{\mathcal{S}}(M)$ is the probability for a single step to land directly in $\mathcal{S}$, as defined in \cref{eq:GammaR_def}. 
\Cref{eq:h_integral_eq} is exact and retains the full discrete structure of the fragmentation kernel. 
Starting from mass $M$, the cascade either succeeds in a single step (probability $\Gamma_{\mathcal{S}}$) or continues to a new mass $M' > M_{\rm cut}$ from which the success probability is $h(M')$.

In the diffusion limit, \cref{eq:h_integral_eq} becomes
\begin{equation}
  \mathcal{L}_{\rm eff}\,h(M)=0,
  \label{eq:h_exact_backward}
\end{equation}
where $\mathcal{L}_{\rm eff}$ is the effective generator governing the bare fragmentation dynamics and contains both local and non-local contributions introduced in \cref{subsec:local_eft,subsec:nonlocal_operators}, respectively. 
\Cref{eq:h_exact_backward} implies that the success probability $h(M)$ satisfies an exact invariance condition under the bare fragmentation dynamics. 
Specifically, away from the absorbing termination and failure regions, the value of $h(M)$ is unchanged on average by an infinitesimal fragmentation step. 
In this sense, $h(M)$ is a globally stationary observable of the stochastic process, with all nontrivial structure arising from the absorbing boundaries. 

Given the boundary conditions in \cref{eq:h_bc}, the action of the non--local generator on $h(M)$ may be written as
\begin{equation}
  \mathcal{L}_{\rm nonloc}\,h(M)
  =
  \Gamma_{\mathcal{S}}(M)\,[1-h(M)]
  -
  \Gamma_{\mathcal{F}}(M)\,h(M),
  \label{eq:Lnonloc_on_h}
\end{equation}
where $\Gamma_{\mathcal{F}}(M)$ encodes non--local transitions that undershoot into the failure set $\mathcal{F}$. 
The backward equation may then be written explicitly as
\begin{equation}
  \mu(M)\,h'(M)
  + \tfrac12 \sigma^2(M)\,h''(M)
  + \Gamma_{\mathcal{S}}(M)\,[1-h(M)]
  - \Gamma_{\mathcal{F}}(M)\,h(M)
  = 0.
  \label{eq:h_backward_explicit}
\end{equation}
Crucially, solutions to \cref{eq:h_backward_explicit} cannot be determined from the local transport coefficients $\mu(M)$ and $\sigma^2(M)$ alone, as the non-local tail operators are essential for enforcing the global termination constraint.

\subsection{Approximate analytic solutions in each EFT}
\label{subsec:h_approx}

\begin{figure}[t]
\centering
\includegraphics[width=0.83\textwidth]{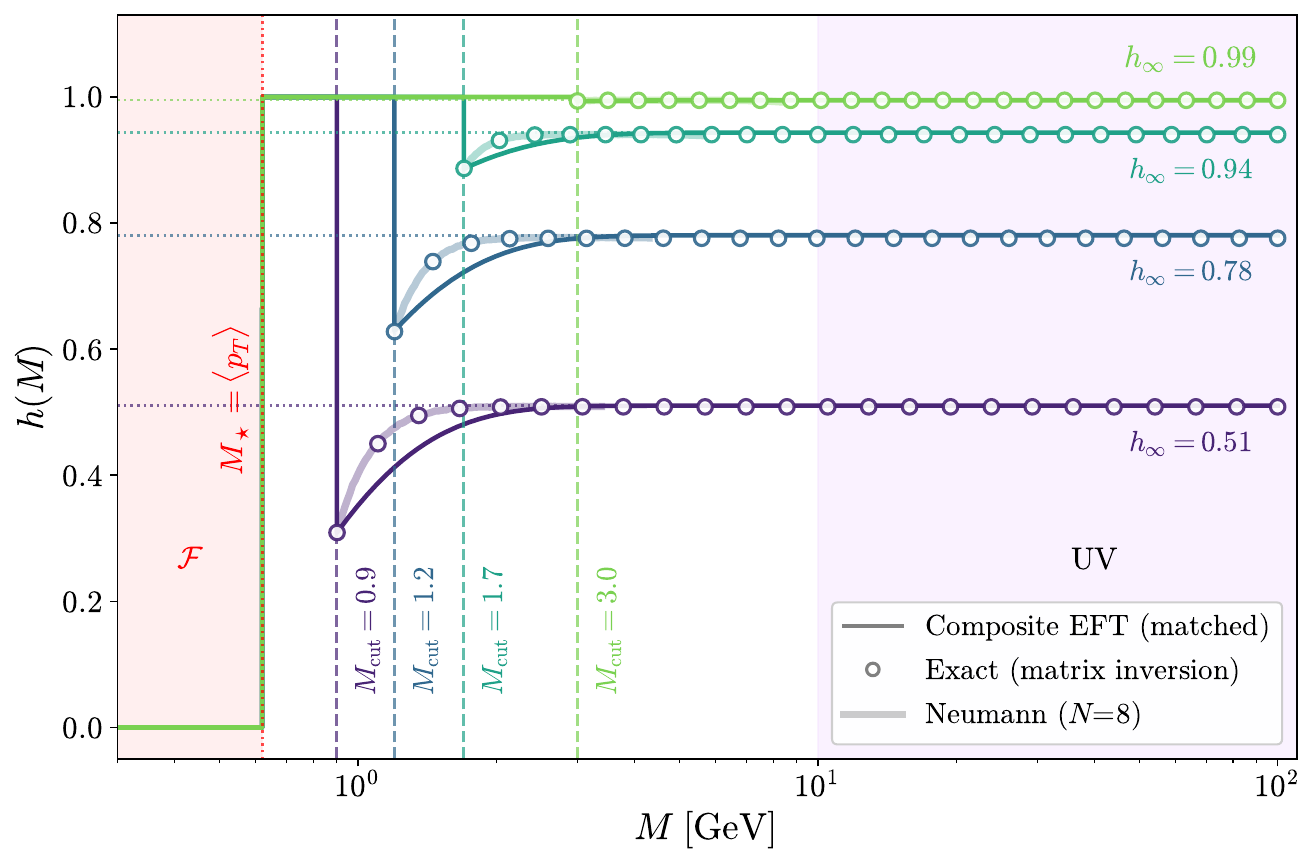}
\caption{Success probability $h(M)$ for cascade termination in $\mathcal{S}$ rather than $\mathcal{F}$, shown for $M_{\rm cut} = 0.9, 1.2, 1.7, 3.0$ GeV with asymptotic plateaus $h_\infty \approx 0.51, 0.78, 0.94, 0.99$ respectively. Solid curves: matched composite EFT interpolating between the boundary, running, and UV approximations. Open circles: exact solution from matrix inversion of the integral equation~\eqref{eq:h_integral_eq}. Broad translucent bands: the truncated Neumann series~\eqref{eq:neumann_series} at order $N=8$, shown in the boundary-layer region within a few multiples of $\ell = 1/|r_-|$ above each $M_{\rm cut}$. The rapid rise near each $M_{\rm cut}$ marks the boundary layer where tail operators enter at leading order. The survival probability formula in \cref{eq:h_inf_complete} achieves $\lesssim 1\%$ accuracy for $h_\infty$ across all cases.}
\label{fig:success_probability}
\end{figure}

The success probability $h(M)$ is a global first-passage observable and, as such, cannot be described exactly by a single local EFT. 
Nevertheless, in each regime of the EFT tower the backward equation admits controlled analytic approximations that illuminate both the structure of the solution and the manner in which conditioning renormalizes the effective dynamics.

The purpose of the analytic solutions presented below is therefore twofold: (i) to make manifest the physical content of each EFT approximation, and (ii) to provide transparent benchmarks against which the exact numerical solution may be compared. 

\subsubsection{UV local EFT -- the asymptotic plateau}

In the ultraviolet regime $M\gg M_{\rm cut}$, order-one fragmentation events are parametrically suppressed and the tail operator contributes only at subleading order. 
To leading order, the backward equation reduces to the homogeneous local diffusion equation
\begin{equation}
  \frac{1}{2}\sigma_{\rm UV}^2(M)\,h''(M)
  + \mu_{\rm UV}(M)\,h'(M)=0,
  \label{eq:h_UV_LO}
\end{equation}
with scale-invariant coefficients $\mu_{\rm UV}(M)=c_1 M$ and $\sigma_{\rm UV}^2(M)=d_1 M^2$.

To solve this equation, we introduce the logarithmic mass variable
\begin{equation}
  x\equiv \ln M,
  \label{eq:x_def}
\end{equation}
and write $h(M) = H(x)$. Under this transformation, derivatives transform as
\begin{equation}
  \frac{dh}{dM} = \frac{1}{M}\,\frac{dH}{dx},
  \qquad
  \frac{d^2h}{dM^2} = \frac{1}{M^2}\left(\frac{d^2H}{dx^2} - \frac{dH}{dx}\right).
  \label{eq:derivative_transform}
\end{equation}
Substituting into \cref{eq:h_UV_LO} yields
\begin{equation}
  H''(x)+\left(\frac{2c_1}{d_1}-1\right)H'(x)=0,
\end{equation}
with general solution
\begin{equation}
  h_{\rm UV}(M)
  =
  A_{\rm UV}+B_{\rm UV}\,M^{\beta},
  \qquad
  \beta \equiv 1-\frac{2c_1}{d_1}.
  \label{eq:h_UV_sol}
\end{equation}
For the Lund kernel one has $c_1<0$ and $d_1>0$, hence $\beta>0$ and the growing mode is excluded by boundedness. 
The UV EFT therefore predicts a constant plateau at large $M$,
\begin{equation}
  h_{\text{UV}}(M) \simeq h_\infty, \qquad h_\infty \equiv \lim_{M\to\infty} h(M),
\end{equation}
with power-suppressed corrections governed by non-local effects.\footnote{%
When the non-local tail operators are reinstated, the backward equation becomes inhomogeneous, and the constant UV plateau acquires controlled, decaying corrections set by the large-$M$ scaling of the tail probabilities $\Gamma_{\mathcal{F},\mathcal{S}}(M)\sim (M_{\rm scale}/M)^{2(a+1)}$. 
These terms generate the asymptotic approach to the plateau observed in the exact solution.}

\subsubsection{Intermediate EFT -- persistent plateau}

At finite but still large string mass, the local drift and diffusion coefficients retain their full dependence on $\lambda \equiv bM^2$,
\begin{equation}
  \mu_{\rm run}(M)=M\,C_1(\lambda),
  \qquad
  \sigma^2_{\rm run}(M)=M^2\,D_1(\lambda),
\end{equation}
with $C_1(\lambda)$ and $D_1(\lambda)$ fixed by moments of the microscopic fragmentation kernel. 
In this regime order-one jumps remain parametrically suppressed, and the backward equation may be treated as local to leading order.

Using the logarithmic transformation introduced in \cref{eq:x_def,eq:derivative_transform}, the homogeneous backward equation becomes
\begin{equation}
  H''(x) + \left[\frac{2C_1(\lambda(x))}{D_1(\lambda(x))}-1\right] H'(x) = 0.
  \label{eq:h_run_x}
\end{equation}
Defining $Y(x) \equiv H'(x)$, this reduces to a first-order flow equation
\begin{equation}
  \frac{dY}{dx} = -\gamma(\lambda(x))\,Y,
  \qquad
  \gamma(\lambda) \equiv \frac{2C_1(\lambda)}{D_1(\lambda)}-1.
  \label{eq:gamma_flow}
\end{equation}
The quantity $\gamma(\lambda)$ plays the role of an anomalous dimension, controlling how the gradient $h'(M)$ runs with the logarithmic mass scale. 
In the UV limit $\lambda \to \infty$, one recovers the fixed-point value $\gamma \to 2c_1/d_1 - 1 = -\beta$.

The general solution can be expressed in closed form as an integral over $\gamma(\lambda)$, but the essential physics is captured by the sign of $\gamma$. 
For the Lund kernel, $\gamma(\lambda) < 0$ throughout the intermediate regime: the ratio $C_1/D_1$ remains negative because the mean step $\langle \Delta M \rangle < 0$ while the variance $\langle (\Delta M)^2 \rangle > 0$. 
The negative anomalous dimension ensures that the growing mode excluded by boundedness in the UV analysis remains excluded at all finite masses. 
The success probability therefore maintains a constant plateau throughout the intermediate regime, with nontrivial $M$-dependence arising only in the boundary layer where non-local effects enter at leading order.

\subsubsection{IR/Boundary EFT -- exponential approach to the plateau}
\label{subsubsection:boundary_eft_h}

In the vicinity of the termination scale $M\simeq M_{\rm cut}$, rare fragmentation events that induce order-one changes in the string mass are no longer parametrically suppressed. 
In this regime, the evolution of the success probability $h(M)$ is governed by a competition between local diffusive transport and direct one-step transitions into either the termination band $\mathcal{S}$ or the failure region $\mathcal{F}$. 
This defines a genuine \emph{boundary layer} in which non-local effects enter at leading order.

\paragraph{Diffusion approximation.}

The backward ODE~\cref{eq:h_backward_explicit} provides a useful starting point for the boundary EFT. 
Freezing the local transport coefficients and tail operators at the termination scale,
\begin{equation}
  \mu(M)\simeq \mu_{\rm cut},
  \qquad
  \sigma^2(M)\simeq \sigma_{\rm cut}^2,
\end{equation}
reduces the backward equation to
\begin{equation}
  \frac{1}{2}\sigma_{\rm cut}^2\,h''(M)
  +\mu_{\rm cut}\,h'(M)
  -\Gamma_{\rm cut}\,h(M)
  +\Gamma_{\mathcal S,{\rm cut}}
  =0,
  \label{eq:h_boundary_LO}
\end{equation}
where
\begin{equation}
  \Gamma_{\mathcal S,{\rm cut}}\equiv\Gamma_{\mathcal S}(M_{\rm cut}),
  \qquad
  \Gamma_{\mathcal F,{\rm cut}}\equiv\Gamma_{\mathcal F}(M_{\rm cut}),
  \qquad
  \Gamma_{\rm cut}
  =
  \Gamma_{\mathcal S,{\rm cut}}+\Gamma_{\mathcal F,{\rm cut}}
\end{equation}
is the total one-step exit rate from the continuation region.

This linear inhomogeneous equation admits a constant particular solution
\begin{equation}
  h_{\rm loc}^{\rm cut}
  =
  \frac{\Gamma_{\mathcal S,{\rm cut}}}{\Gamma_{\rm cut}},
\end{equation}
representing the local branching fraction into $\mathcal{S}$ via a single tail event at $M_{\rm cut}$. 
The general solution is
\begin{equation}
  h_{\rm bdry}(M)
  =
  h_{\rm loc}^{\rm cut}
  +
  A\,e^{r_+(M-M_{\rm cut})}
  +
  B\,e^{r_-(M-M_{\rm cut})},
\end{equation}
with characteristic exponents
\begin{equation}
  r_\pm
  =
  \frac{-\mu_{\rm cut}\pm\sqrt{\mu_{\rm cut}^2+2\sigma_{\rm cut}^2\Gamma_{\rm cut}}}
       {\sigma_{\rm cut}^2}.
\end{equation}
Since $r_+>0$, boundedness requires $A=0$, leaving only the decaying mode,
\begin{equation}
  h_{\rm bdry}(M)
  =
  h_{\rm loc}^{\rm cut}
  +
  B\,e^{r_-(M-M_{\rm cut})},
  \qquad r_- < 0.
  \label{eq:h_bdy_solution}
\end{equation}

The boundary EFT captures the vicinity of $M_{\rm cut}$ where tail operators enter at leading order, but the integration constant $B$ remains undetermined. 
The constant $h_{\rm loc}^{\rm cut}$ represents only the \emph{local} branching fraction at the cut---the probability of landing in $\mathcal{S}$ rather than $\mathcal{F}$ given a single tail event at $M_{\rm cut}$---not the global success probability $h_\infty$ from large $M$. 
Connecting the boundary solution to the asymptotic plateau and fixing $B$ requires matching to the outer EFT, developed in \cref{subsec:h_matching}.

\paragraph{Exact boundary theory.}

The integral equation~\cref{eq:h_integral_eq} admits a natural perturbative solution. 
Defining the continuation probability
\begin{equation}
  P_{\rm cont}(M) \equiv \int_{C} K(M \to M')\,dM',
  \label{eq:P_cont}
\end{equation}
the probability that a fragmentation step starting at $M$ remains in the continuation region $C \equiv (M_{\rm cut}, \infty)$, and the continuation operator
\begin{equation}
  (T f)(M) \equiv \int_{C} K(M \to M')\, f(M')\,dM',
  \label{eq:T_operator}
\end{equation}
the integral equation becomes $h = \Gamma_{\mathcal{S}} + Th$, which rearranges to
\begin{equation}
  (I - T)\,h = \Gamma_{\mathcal{S}}.
  \label{eq:h_linear_operator}
\end{equation}
The operator $T$ is \emph{substochastic}: at each mass $M$, the continuation probability satisfies $P_{\rm cont}(M) < 1$, since every fragmentation step has a nonzero probability of exiting $C$ into either $\mathcal{S}$ or $\mathcal{F}$. 
The substochastic property ensures that the operator norm of $T$ is strictly less than unity, so $(I - T)$ is invertible and the formal solution of \cref{eq:h_linear_operator} may be expanded as a geometric series,
\begin{equation}
  h = (I - T)^{-1}\,\Gamma_{\mathcal{S}}
  = \sum_{n=0}^{\infty} T^n\, \Gamma_{\mathcal{S}}.
  \label{eq:neumann_series}
\end{equation}
This power series expansion in the continuation operator is known as a Neumann series. 
Each term admits a direct probabilistic interpretation where $T^n \Gamma_{\mathcal{S}}(M)$ is the probability for a cascade starting at mass $M$ to make exactly $n$ continuation steps within $C$ before exiting into $\mathcal{S}$. 
The leading term $\Gamma_{\mathcal{S}}(M)$ captures direct exit without any intervening continuation steps, while higher terms progressively incorporate the effect of multiple re-entries into $C$ before eventual success.

Convergence is controlled by $P_{\rm cont}(M)$, which serves as the perturbative expansion parameter. 
Near $M_{\rm cut}$, where $P_{\rm cont} \ll 1$, convergence is rapid and the leading term already captures the dominant contribution and each successive application of $T$ is suppressed by the small probability of continued fragmentation within $C$. 
The full solution retains the complete discrete structure of the fragmentation kernel without moment truncation, providing the exact inner solution to the boundary theory. 
The practical construction of $T$ and $\Gamma_{\mathcal{S}}$ from the microscopic kernel, and the convergence properties of the numerical solution, are described in \cref{subsubsec:integral_eq_solve}.

\subsection{Matching the EFT tower}
\label{subsec:h_matching}

In effective field theory, descriptions valid at different scales are connected through matching: physical observables computed in overlapping regimes must agree. 
The present construction follows this logic with one essential difference: the Wilson coefficients $\{\mu(M), \sigma^2(M), \Gamma_{\mathcal{F}}(M), \Gamma_{\mathcal{S}}(M)\}$ are already determined by the microscopic Lund kernel. 
What matching determines are the \emph{integration constants} that parameterize solutions of the backward equation $\mathcal{L}_{\rm eff}^{(i)} h^{(i)}(M) = 0$ in each regime.

In general, at a matching point $M_{\rm match}$ in the overlap region between two EFT descriptions, we impose continuity of the success probability and its gradient,
\begin{equation}
  h^{(i)}(M_{\rm match}) = h^{(j)}(M_{\rm match}),
  \qquad
  \partial_{M} h^{(i)}(M_{\rm match}) = \partial_{M} h^{(j)}(M_{\rm match}).
  \label{eq:h_matching_conditions}
\end{equation}
Derivative matching is imposed because $h'(M)$ can enter physical observables. 
For instance, the conditioned dynamics of \cref{sec:doob_h} depend on $\partial_M \ln h$, which would be discontinuous without derivative matching.

\subsubsection{The boundary--UV matching problem}

Connecting the boundary-layer solution~\cref{eq:h_bdy_solution} to the asymptotic plateau requires additional care. 
As $M$ increases beyond the boundary layer, the decaying mode is exponentially suppressed and the inner solution asymptotes to $h_{\rm loc}^{\rm cut}$ rather than $h_\infty$. 
Matching at a finite scale $M_{\rm match}$ by imposing $h_{\rm bdry}(M_{\rm match}) = h_\infty$ would force $B \propto e^{-r_-(M_{\rm match} - M_{\rm cut})}$, producing an exponentially large overshoot at $M_{\rm cut}$. 
The gap $h_\infty - h_{\rm loc}^{\rm cut}$ is physical, arising because freezing the tail operators at $M_{\rm cut}$ removes the running of $\Gamma_{\mathcal{S}}(M)/\Gamma(M)$ that drives $h$ toward the plateau, so $h_\infty$ must be computed independently (\cref{subsubsec:h_inf_derivation}).

Given $h_\infty$ (see \cref{subsubsec:h_inf_derivation}), we construct a composite approximation via standard matched asymptotics, $h_{\rm comp} = h_{\rm inner} + h_{\rm outer} - h_{\rm overlap}$, with outer solution $h_{\rm outer} = h_\infty$ and overlap value $h_{\rm overlap} = h_{\rm loc}^{\rm cut}$,
\begin{equation}
  h_{\rm comp}(M)
  =
  h_\infty
  + (h_{\rm loc}^{\rm cut} - h_\infty)\,e^{r_-(M-M_{\rm cut})},
  \label{eq:bdy_matched_form}
\end{equation}
interpolating between the local branching fraction at $M_{\rm cut}$ and the many-step survival average $h_\infty$ over a characteristic boundary-layer width $\ell \equiv 1/|r_-|$.

Compared to the exact integral equation~\cref{eq:h_integral_eq}, the composite solution exhibits a residual ${\sim}\,8\%$ discrepancy in the boundary layer (\cref{fig:success_probability}). 
This reflects a structural limitation of the diffusion approximation. 
Near $M_{\rm cut}$, the total exit rate $\Gamma(M)$ approaches unity and the continuation probability drops to ${\sim}\,10^{-3}$, so essentially every fragmentation step exits the continuation region. 
The boundary layer is therefore not a regime where diffusion competes with exits, but a crossover from many-step stochastic evolution to immediate single-step termination---a transition the drift-diffusion ODE cannot capture, since its transport terms become irrelevant precisely where the boundary condition is imposed. 
The composite~\cref{eq:bdy_matched_form} succeeds because it bridges two independently determined limits without relying on the ODE in the transition zone.

For higher precision, the exact inner solution---obtained by solving \cref{eq:h_integral_eq} or equivalently summing the Neumann series~\cref{eq:neumann_series} to convergence---retains the discrete jump structure and provides genuine matched asymptotics. 
Near $M_{\rm cut}$, the Neumann series converges rapidly because $P_{\rm cont} \ll 1$. 
As $M$ increases, $P_{\rm cont}(M) \to 1$ and more terms are needed, but simultaneously $h(M) \to h_\infty$. 
An overlap region exists from ${\sim}\,\ell$ to ${\sim}\,5$--$7\,\ell$ above $M_{\rm cut}$, in which the truncated Neumann series at modest order ($N \approx 8$) has converged to better than $10^{-4}$ while the exact solution has approached $h_\infty$ to within a percent.

\subsubsection{Determining the asymptotic plateau}
\label{subsubsec:h_inf_derivation}

A cascade starting at large $M$ drifts downward until a tail event terminates the evolution, with the outcome at the moment of exit determined by the local branching fraction $h_{\rm loc}(M) \equiv \Gamma_{\mathcal{S}}(M)/\Gamma(M)$.
The asymptotic success probability receives contributions from cascades that exit above the boundary and those that survive to it. 
The former contribute
\begin{equation}
  h_\infty^{\rm bulk} = \int_{M_{\rm cut}}^\infty h_{\rm loc}(M) \, P_{\rm term}(M) \, dM,
  \label{eq:h_inf_bulk}
\end{equation}
where $P_{\rm term}(M)$ is the probability density for a cascade to first exit at mass $M$. 
The termination density follows from a survival-probability argument. 
The exit probability per unit mass traversed is $\Gamma(M)/|\mu(M)|$, so the probability of surviving to mass $M$ without exiting earlier is
\begin{equation}
  P_{\rm surv}(M) = \exp\!\left( -\int_M^\infty \frac{\Gamma(M')}{|\mu(M')|} \, dM' \right),
\end{equation}
and the termination density is
\begin{equation}
  P_{\rm term}(M) = \frac{\Gamma(M)}{|\mu(M)|} \, P_{\rm surv}(M).
  \label{eq:P_term}
\end{equation}
A fraction $P_{\rm surv}(M_{\rm cut}) \approx 0.2$\,--\,$0.3$ of cascades survive to $M_{\rm cut}$ without exiting earlier, contributing
\begin{equation}
  h_\infty^{\rm bdry} = h_{\rm loc}(M_{\rm cut}) \, P_{\rm surv}(M_{\rm cut}).
  \label{eq:h_inf_bdry}
\end{equation} 
The total asymptotic success probability is then
\begin{equation}
  h_\infty = h_\infty^{\rm bulk} + h_\infty^{\rm bdry},
  \label{eq:h_inf_complete}
\end{equation}
which achieves 2--3\% agreement with Monte Carlo estimates.

\begin{figure}[t!]
\centering
\includegraphics[width=1.0\textwidth]{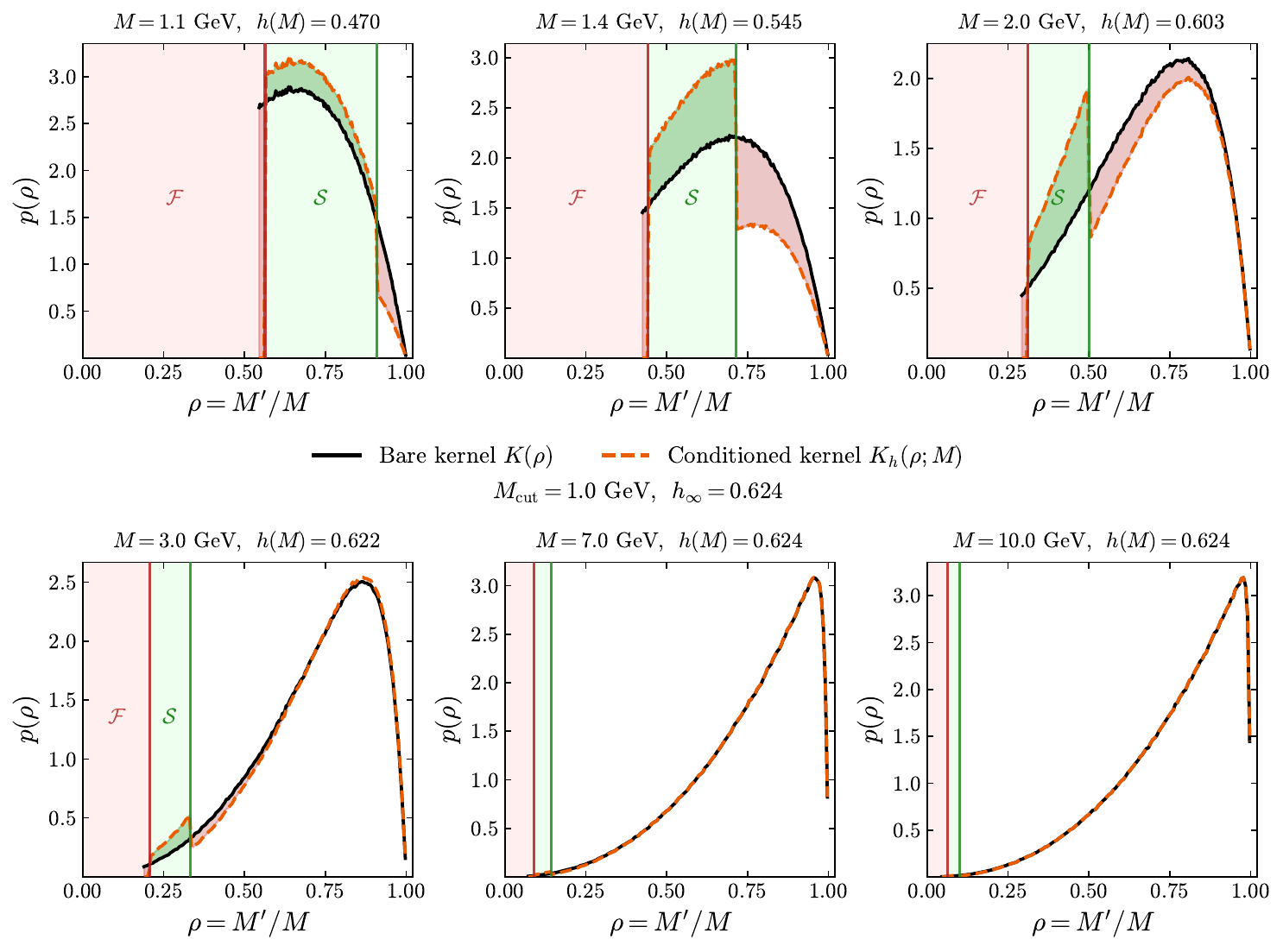}
\caption{Doob $h$-transformed fragmentation kernel $K_h(M \to M') = K_{\rm bare}(M \to M') \cdot h(M')/h(M)$ for $M_{\rm cut} = 1.0$ GeV ($h_\infty = 0.62$). Each panel compares the bare kernel (solid black) to the conditioned kernel (dashed) at $M = 1.1, 1.4, 2.0, 3.0, 7.0, 10.0$ GeV. Near $M_{\rm cut}$, the $h$-transform suppresses large downward jumps that risk undershooting into $\mathcal{F}$ while enhancing transitions that land safely in $\mathcal{S}$. At large $M$, where $h(M') \approx h(M) \approx h_\infty$, the conditioning effect vanishes and $K_h \to K_{\rm bare}$.}
\label{fig:effective kernel}
\end{figure}

\subsection{Numerical validation}
\label{subsec:eft_validation}

All numerical results in this paper use Monash tune parameters ($a = 0.68$, $b = 0.98$~GeV$^{-2}$)~\cite{Skands:2014pea} in the chiral limit, giving $M_\star = 0.62$~GeV. 
The ``Monte Carlo'' reference values quoted throughout are obtained from two independent procedures described below: direct simulation of bare cascades, and numerical solution of the integral equation~\cref{eq:h_integral_eq}.

\subsubsection{Bare cascade simulation}

The bare fragmentation cascade is simulated as a discrete Markov chain in which, at each step, the scaled variables $u$ and $v$ are sampled independently from the factorized distributions~\cref{eq:factorized_distributions} via inverse-CDF sampling, yielding a mass ratio $\rho \equiv M'/M = \sqrt{(1-u)(1-v)}$ that updates the string mass through $M_{n+1} = \rho\, M_n$. 
The cascade terminates when $M_n \leq M_{\rm cut}$ and is classified as successful ($M_n \in \mathcal{S}$) or failed ($M_n \in \mathcal{F}$). 
The empirical success rate over a large ensemble of cascades initiated at $M_0$ provides a direct Monte Carlo estimate of $h_\infty$. 
The kernel densities in~\cref{fig:effective kernel} are constructed by histogramming sampled mass ratios $\rho$ at fixed $M$ and reweighting each entry by $h(\rho\, M)/h(M)$ to obtain the conditioned distribution. 
For the conditioned dynamics in~\cref{fig:conditioned_dynamics}, the $h$-transformed kernel is implemented via rejection sampling from $\mathcal{K}_{\rm bare}$ with acceptance probability $h(M')$, as described in~\cref{sec:practical}.

\subsubsection{Solving the integral equation}
\label{subsubsec:integral_eq_solve}

The integral equation~\cref{eq:h_integral_eq} provides an exact determination of $h(M)$ that retains the full discrete structure of the fragmentation kernel.  
In practice, the operator equation~\cref{eq:h_linear_operator} is solved by direct matrix inversion, which we now describe.

The continuation region $C = (M_{\rm cut}, M_0)$ is discretized on a logarithmic grid $\{M_1, M_2, \ldots, M_N\}$ with $M_1$ just above $M_{\rm cut}$, and $h(M)$ is replaced with a discrete vector $\mathbf{h} = (h_1, \ldots, h_N)$. 
At each grid point $M_i$, a large number of mass ratios $\rho$ are sampled from the bare kernel and the daughter masses $M' = \rho\, M_i$ are classified into three channels. 
Daughters landing in the success band contribute to the direct termination probability $\Gamma_{\mathcal{S},i} \equiv \mathbb{P}(\rho\, M_i \in \mathcal{S})$. Daughters falling in the failure region are discarded, since $h = 0$ on $\mathcal{F}$. 
Daughters remaining in the continuation region $M' > M_{\rm cut}$ are binned to estimate the transition matrix $T_{ij} \equiv \mathbb{P}(\rho\, M_i \in \mathrm{bin}_j)$, whose row sums give the continuation probability $P_{{\rm cont},i} = \sum_j T_{ij}$. 
With these inputs, the discretized form of \cref{eq:h_linear_operator}, $(I - T)\,\mathbf{h} = \boldsymbol{\Gamma}_{\mathcal{S}}$, is solved in a single pass by LU decomposition, yielding $h_i$ at every grid point simultaneously.

The substochastic property $P_{{\rm cont},i} < 1$ guarantees convergence, whether the solution is obtained by matrix inversion or by iterating the Neumann series~\cref{eq:neumann_series}. 
Near $M_{\rm cut}$, the continuation probability drops to $P_{\rm cont} \sim 10^{-3}$ and the series converges in a few terms; at larger masses where $P_{\rm cont} \to 1$, more terms are required. 
In practice, $N \approx 8$ Neumann iterations suffice for $< 10^{-4}$ accuracy throughout the boundary layer, while the matrix solve achieves machine precision in a single pass at cost $\mathcal{O}(N^3)$.

\subsubsection{Comparison of approximations}

\Cref{fig:success_probability} compares the matched composite EFT approximation~\cref{eq:bdy_matched_form} with asymptotic plateau $h_\infty$ from~\cref{eq:h_inf_complete} against the exact integral equation solution and direct Monte Carlo simulation. The integral equation achieves sub-percent agreement with Monte Carlo across all mass scales; the matched boundary EFT with survival-probability estimate of $h_\infty$ achieves $\sim$8\% accuracy at worst in the boundary layer, improving to 2--3\% in the asymptotic regime, as discussed in~\cref{subsec:h_matching}.

The success probability $h(M)$ constructed in this section provides a testing ground for comparing the EFT tower against exact benchmarks and reveals the limits of validity of each approximation. 
It also serves as the renormalization function for the Doob $h$-transform developed in \cref{sec:stoch_dynamics}. 
Given $h(M)$, we can now construct the renormalized generator and extract its physical interpretation.

\section{Conditioning and local renormalization}
\label{sec:doob_h}

Having computed the success probability $h(M)$ across all mass scales, we now construct the conditioned dynamics explicitly. 
Applying the Doob $h$-transform from \cref{sec:stoch_dynamics}, the renormalized generator $\mathcal{L}_h = h^{-1}\mathcal{L}_{\text{eff}}h$ takes the form
\begin{equation}
  \mathcal{L}_h \phi = \mu_h(M) \partial_M \phi + \tfrac{1}{2} \sigma^2(M) \partial_M^2 \phi + \mathcal{L}_{h,\text{tail}} \phi,
  \label{eq:renormalized_generator_hadronization}
\end{equation}
with \emph{renormalized drift}
\begin{equation}
  \mu_h(M) = \mu(M) + \sigma^2(M) \, \partial_M \ln h(M),
  \label{eq:renormalized_drift_hadronization}
\end{equation}
where the additional term follows from expanding the conjugation and using the backward equation $\mathcal{L}_{\text{eff}} h = 0$. 
The diffusion coefficient $\sigma^2(M)$ is unchanged by conditioning, while tail operators acquire modified jump distributions through $h(M')/h(M)$.

The correction to the drift admits a natural interpretation as an emergent force derived from an effective potential $V_{\text{eff}}(M) \equiv -\ln h(M)$:
\begin{equation}
  F(M) \equiv \sigma^2(M) \, \partial_M \ln h(M) = -\sigma^2(M) \, \partial_M V_{\text{eff}}(M).
  \label{eq:emergent_force}
\end{equation}
This force biases the string mass evolution toward configurations with higher viability, encoding in a local and EFT-compatible manner the effect of global energy-momentum conservation on the stochastic cascade. 
We refer to this mechanism as \emph{budget awareness}. 
The sign and magnitude of $F(M)$ are determined by the gradient of the success probability:
\begin{itemize}
\item \textbf{Positive gradient} ($h'(M) > 0$): The force $F(M) > 0$ opposes the bare drift, slowing the approach to the termination region. 
This occurs when increasing $M$ improves the chances of successful completion, signaling that the cascade has ``overshot'' and should be pulled back.

\item \textbf{Negative gradient} ($h'(M) < 0$): The force $F(M) < 0$ reinforces the bare drift, accelerating the approach to $\mathcal{S}$. 
This occurs in boundary layers where direct transitions into $\mathcal{S}$ via tail operators become favorable, pulling the cascade toward successful termination.

\item \textbf{Plateau regions} ($h'(M) \approx 0$): The force vanishes, and the renormalized dynamics reduce to the bare evolution. This occurs far from boundaries where success probability is approximately constant. 
\end{itemize}
The diffusion coefficient $\sigma^2(M)$ acts as an amplification factor, ensuring that the emergent force is most significant near kinematic boundaries where stochastic fluctuations are comparable to the remaining phase space.

\begin{figure}[t!]
\centering
\includegraphics[width=1.0\textwidth]{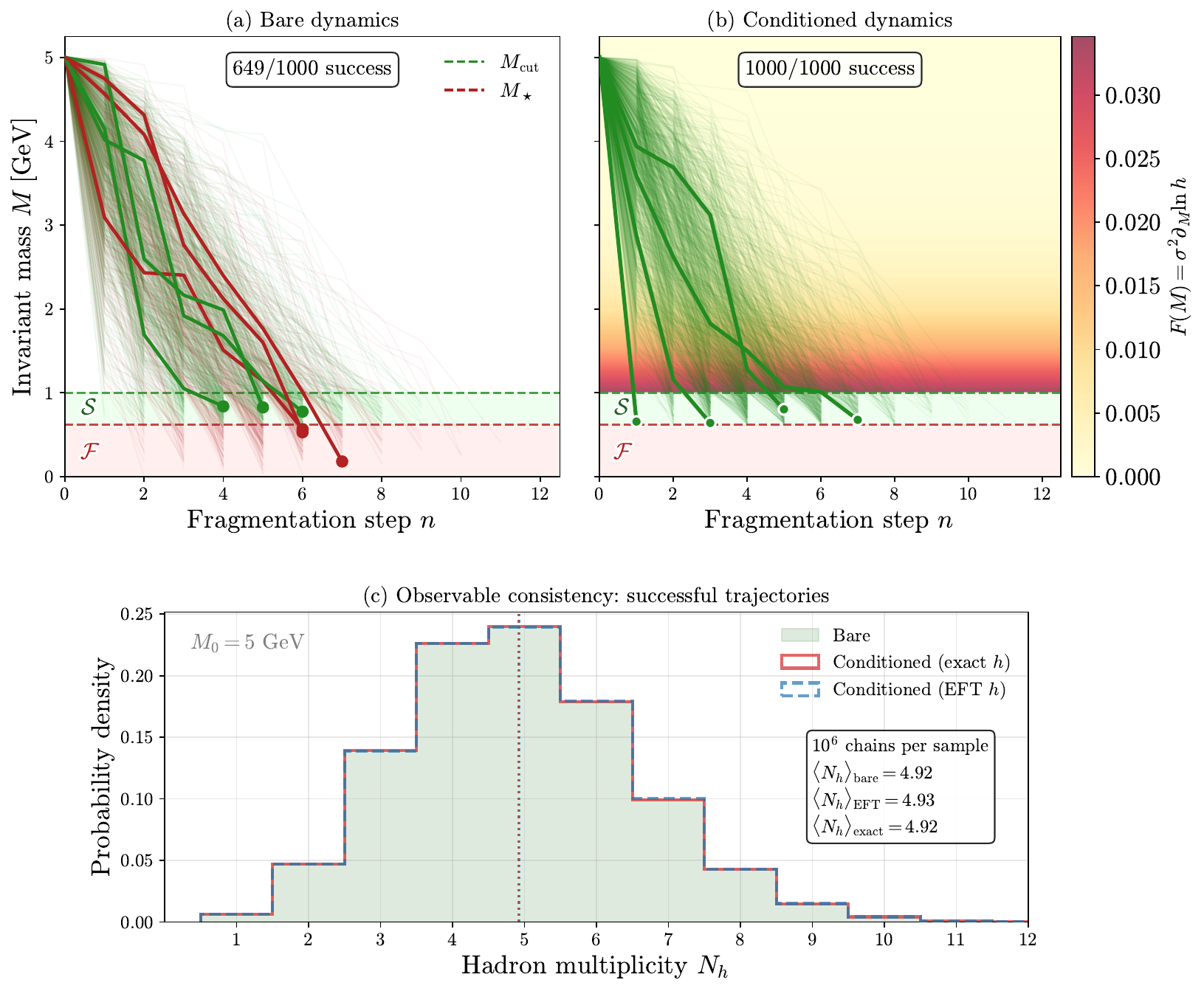}
\caption{Comparison of bare and conditioned fragmentation dynamics for $M_0 = 5.0$ GeV and $M_{\rm cut} = 1.0$ GeV. \textbf{(a)} Sample trajectories under the bare kernel; 649/1000 reach the termination band $\mathcal{S} = [M_\star, M_{\rm cut}]$ (green) while the remainder undershoot into $\mathcal{F} = (0, M_\star)$ (red), with $M_\star = 0.62$ GeV. \textbf{(b)} Trajectories under the $h$-transformed kernel $K_h$, where the background color map encodes the emergent force $F(M) = \sigma^2 \partial_M \ln h$; all 1000/1000 conditioned trajectories terminate successfully by construction. \textbf{(c)} Hadron multiplicity distributions from $10^6$ successful trajectories each, comparing bare dynamics (shaded), conditioned sampling with exact $h$ (solid), and conditioned sampling with EFT $h$ (dashed). The three distributions are statistically indistinguishable ($\langle N_h \rangle \approx 4.92$), confirming that the $h$-transform preserves physical observables while eliminating failed trajectories.}
\label{fig:conditioned_dynamics}
\end{figure}

This structure clarifies the relationship between the bare Lund model and physical 
hadronization. 
The Lund kernel provides a local, factorized \emph{proposal distribution} encoding the fundamental string-breaking dynamics, while the physical ensemble is obtained by conditioning on successful termination. 
The function $h(M)$ acts as a non-equilibrium analogue of the Boltzmann weight, propagating a late-time constraint backward through the cascade via local trajectory reweighting. 
Unlike equilibrium statistical mechanics where potentials are imposed by external fields, here the effective potential is self-generated by the requirement of consistent cascade completion.

\subsection{EFT structure of the renormalized dynamics}
\label{subsec:eft_renormalization}

The emergent force $F(M) = \sigma^2(M) \partial_M \ln h(M)$ inherits the EFT tower structure from the success probability constructed in \cref{sec:h_global}. 
In each kinematic regime, the force admits a transparent physical interpretation.

\subsubsection{UV regime -- dormant foresight}
Far from termination, the success probability approaches a constant plateau, $h(M) \to h_\infty$, reflecting the fact that a high-mass string has ample phase space to fragment successfully regardless of small variations in its current state. The effective potential $V_{\text{eff}} = -\ln h$ is nearly flat, so no force is required: the cascade evolves according to the bare Lund dynamics with power-suppressed corrections,
\begin{equation}
  F_{\rm UV}(M) = \mathcal{O}(M^{1-\alpha}), \quad \alpha = \left| \frac{2c_1}{d_1} - 1 \right| > 0.
\end{equation}

\subsubsection{Running regime -- non-local suppression}
In the intermediate regime, the success probability remains approximately constant, so the local drift correction is still suppressed. 
However, conditioning acts non-trivially through the tail operators. 
For rare large jumps that would land in or near the boundary layer, the reweighting factor $h(M')/h(M)$ departs from unity: jumps overshooting toward the failure region $\mathcal{F}$ are suppressed or eliminated, even though the local diffusive dynamics remain essentially bare. 
The constraint thus ``reaches back'' through the tails, filtering out catastrophic trajectories before the cascade enters the boundary layer. 
This non-local awareness coexists with running transport coefficients $\mu(M)$ and $\sigma^2(M)$ that encode the bare fragmentation dynamics.

\subsubsection{Boundary regime -- active braking}
Near the termination threshold, success versus failure becomes highly sensitive to the current mass. 
The success probability varies rapidly,
\begin{equation}
  h_{\rm bdry}(M) = h_{\rm loc} + B_{\rm bdry} \, e^{r_-(M - M_{\rm cut})}, \quad r_- < 0,
\end{equation}
and the effective potential steepens accordingly. 
The emergent force becomes order-one:
\begin{equation}
  F_{\rm bdry}(M) 
  = \sigma^2_{\rm bdry} \, r_- \, \frac{B_{\rm bdry} \, e^{r_-(M - M_{\rm cut})}}{h_{\rm loc} + B_{\rm bdry} \, e^{r_-(M - M_{\rm cut})}}.
\end{equation}
Since matching to the running EFT requires $h(M)$ to rise from $h_{\rm loc}$ toward $h_\infty$ as $M$ increases, the gradient $h'(M) > 0$ is positive throughout the boundary layer. 
The resulting positive force opposes the bare drift, effectively braking the cascade and providing additional fragmentation attempts near the termination threshold (see \cref{fig:conditioned_dynamics}). 
This increases the probability of landing in the success band $\mathcal{S}$ rather than overshooting into the failure region $\mathcal{F}$.

\subsubsection{Tail operators}
The non-local tail operators also renormalize under the $h$-transform. 
Jumps into the termination region $\mathcal{S}$ acquire a factor $1/h(M)$, amplifying transitions from low-success-probability configurations, while jumps into the failure region $\mathcal{F}$ are eliminated by the factor $h(M_{\mathcal{F}}) = 0$. 
This ensures that all components of the generator---drift, diffusion, and rare large jumps---are biased toward successful termination consistently across the EFT tower.

The conditioning mechanism thus operates through two channels: a local drift correction that becomes significant only in the boundary layer, and non-local tail modifications that filter large jumps throughout the cascade. 
Both effects vanish in the deep UV and strengthen as the termination threshold approaches, ensuring that all components of the generator are biased toward successful termination consistently across the EFT tower.

\section{Practical implementation and computational cost}\label{sec:practical}

\begin{figure}[t!]
\centering
\includegraphics[width=0.65\textwidth]{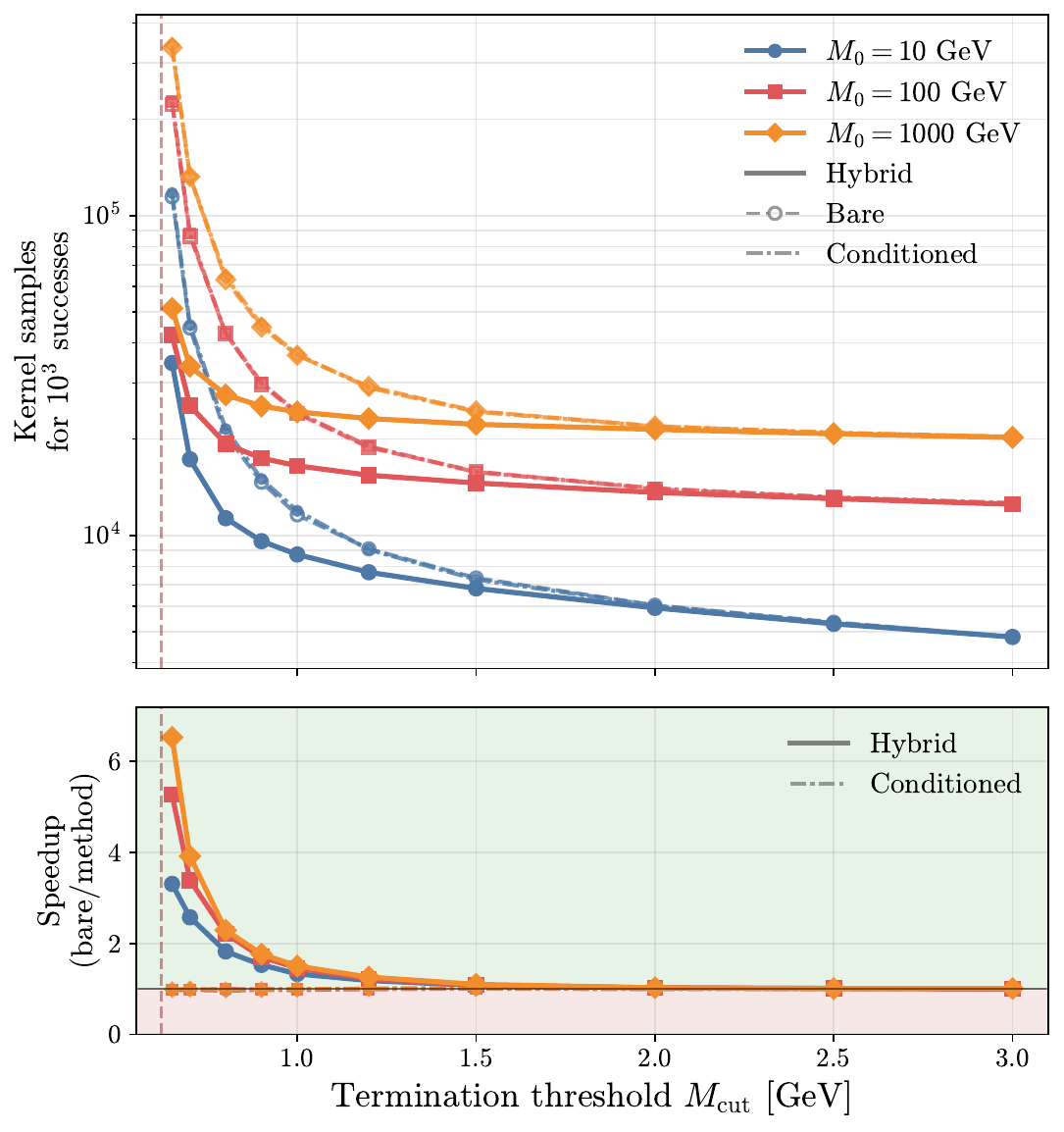}
\caption{Computational cost of generating $10^3$ successful cascades as a function of the termination threshold $M_{\rm cut}$, for initial masses $M_0 = 10, 100, 1000$ GeV. \textbf{(a)} Total kernel samples (including rejection overhead) for three strategies: bare dynamics with rejection filtering (dashed), fully conditioned dynamics (dash-dot), and a hybrid scheme that switches from bare to conditioned evolution at $M_{\rm threshold} = M_{\rm cut} + 3/|r_-|$ (solid). The vertical dashed line marks $M_\star = 0.62$ GeV. \textbf{(b)} Speedup ratio (bare cost divided by method cost); green shading indicates where the conditioned or hybrid sampler outperforms bare rejection filtering. Near $M_\star$, where the bare success rate vanishes, the hybrid sampler achieves up to $\sim\! 6\times$ speedup. As $M_{\rm cut}$ increases the bare success rate approaches unity and the advantage of conditioning diminishes.}
\label{fig:computational_cost}
\end{figure}

The Doob $h$-transform provides an exact rewriting of the conditioned ensemble, in which sampling from $\mathcal{K}_h$ produces trajectories distributed according to $\mathbb{P}_{\text{bare}}(\cdot \mid \mathcal{C})$ without post-hoc filtering. 
It is natural to ask whether this formal exactness translates into a computational advantage over the naive strategy of generating trajectories from $\mathcal{K}_{\text{bare}}$ and discarding those that fail to terminate in $\mathcal{S}$.

The simplest approach is bare generate-and-test, in which complete trajectories are sampled from $\mathcal{K}_{\text{bare}}$ and discarded unless they terminate in $\mathcal{S}$. 
Each trial trajectory of length $L$ (equivalently, hadron multiplicity $N_h$) succeeds with probability $h(M_0) \approx h_\infty$, so the expected number of full trajectories required to obtain one success is $1/h_\infty$, giving a total cost
\begin{equation}
  C_{\text{bare}}
  = \frac{L}{h_\infty}
  = L + L\!\left(\frac{1}{h_\infty} - 1\right),
  \label{eq:cost_bare}
\end{equation}
where the first term represents the irreducible work of a successful cascade and the second the overhead from discarded trajectories. 
The cascade length scales logarithmically, $L \sim \ln(M_0^2/M_{\rm cut}^2)$, so this overhead grows extensively with the initial string mass.

Consider now an implementation of the $h$-transformed dynamics that uses $\mathcal{K}_{\text{bare}}$ as a proposal distribution for $\mathcal{K}_h$. 
At each step, a candidate $M \to M'$ is drawn from $\mathcal{K}_{\text{bare}}$ and accepted with probability proportional to $h(M')/h(M)$. 
Since $h \leq 1$, standard rejection sampling requires on average $1/h(M)$ proposals per accepted transition, so the total cost can be decomposed as
\begin{equation}
  C_{\text{cond}}
  = \sum_{i=1}^{L}\frac{1}{h(M_i)}
  = L + \sum_{i=1}^{L}\!\left(\frac{1}{h(M_i)} - 1\right)
  \;\approx\; L + L\!\left(\frac{1}{h_\infty} - 1\right),
  \label{eq:cost_cond}
\end{equation}
where the approximation holds because the vast majority of steps occur at $M \gg M_{\rm cut}$ where $h(M) \approx h_\infty$. 
The two costs in \cref{eq:cost_bare,eq:cost_cond} are identical. 
Both decompose into the same signal (the irreducible work $L$) and the same noise (the rejection overhead $L(1/h_\infty - 1)$). 
The only difference is whether rejection is applied globally, discarding entire trajectories, or locally, discarding individual transitions. 
When $h_\infty$ is close to unity both methods are efficient; when $h_\infty \ll 1$ both are equally wasteful.

However, the EFT structure developed in the preceding sections reveals that the $h$-transform modifies the bare dynamics only in the boundary layer near $M_{\rm cut}$. 
In the UV and running regimes, $h(M) \approx h_\infty$ and the emergent force $F(M) = \sigma^2 \partial_M \ln h$ is power-suppressed, so $\mathcal{K}_h$ is effectively indistinguishable from $\mathcal{K}_{\text{bare}}$. 
This motivates a hybrid sampling scheme that applies the $h$-transform selectively. 
Define a threshold mass
\begin{equation}
  M_{\rm th} = M_{\rm cut} + n_\sigma / |r_-|,
  \label{eq:M_threshold}
\end{equation}
where $n_\sigma$ controls the number of boundary-layer widths included in the conditioned region (in practice $n_\sigma = 3$ suffices). 
The hybrid sampler evolves the cascade using $\mathcal{K}_{\text{bare}}$ for $M > M_{\rm th}$ with no rejection overhead, and switches to $\mathcal{K}_h$ only when $M$ enters the boundary layer. 
Since the boundary-layer width $1/|r_-|$ is independent of $M_0$, the conditioned-phase cost is $\mathcal{O}(1)$ regardless of the initial string mass, and the total cost reduces to
\begin{equation}
  C_{\text{hybrid}} = L + \mathcal{O}(1),
  \label{eq:cost_hybrid}
\end{equation}
compared with $L/h_\infty$ for both the bare and fully conditioned samplers. 
The hybrid scheme eliminates the extensive rejection overhead entirely, achieving a speedup of $1/h_\infty$ that becomes significant as $M_{\rm cut} \to M_\star$ and $h_\infty \to 0$. 
This behavior is confirmed numerically in \cref{fig:computational_cost}, where the hybrid sampler achieves rapid efficiency gains near $M_\star$.

The hybrid scheme introduces a controlled approximation, since trajectories above $M_{\rm th}$ are drawn from $\mathcal{K}_{\text{bare}}$ rather than $\mathcal{K}_h$. 
The resulting bias is determined by the deviation of $h(M)/h_\infty$ from unity, which is exponentially suppressed above the boundary layer,
\begin{equation}
  \frac{h(M)}{h_\infty}
  = 1 + \mathcal{O}\!\left(e^{r_-(M - M_{\rm cut})}\right),
  \quad M > M_{\rm th},
  \label{eq:bias_suppression}
\end{equation}
with residual bias of order $e^{-n_\sigma} \approx 5\%$ at the switching point. 
When exact results are required, this bias can be removed entirely through importance weighting. 
The likelihood ratio between the exact conditioned measure and the hybrid measure, for a trajectory that crosses $M_{\rm th}$ at step $\tau$, telescopes to
\begin{equation}
  w = \prod_{i=0}^{\tau-1}
      \frac{\mathcal{K}_h(M_i, M_{i+1})}
           {\mathcal{K}_{\text{bare}}(M_i, M_{i+1})}
    = \frac{h(M_\tau)}{h(M_0)}.
  \label{eq:importance_weight}
\end{equation}
Since $h(M_\tau) \approx h_\infty \approx h(M_0)$ for $M_\tau, M_0 \gg M_{\rm cut}$, the weight $w \approx 1$ for the vast majority of trajectories and the variance of the importance-weighted estimator is negligible. 
This provides an exact correction at no additional cost beyond two evaluations of $h$, which is already available from the EFT or the integral equation~\eqref{eq:h_integral_eq}.

The hybrid scheme thus combines the formal advantages of the $h$-transform, namely guaranteed success and exact observables after reweighting, with the computational efficiency of bare sampling in the UV. 
The localization of conditioning overhead to the boundary layer is a direct consequence of the EFT structure developed in the preceding sections. 
The emergent force is an infrared effect with finite, $M_0$-independent cost, and applying it only where it is dynamically active yields a parametrically more efficient sampler.

\section{Conclusions and outlook}\label{sec:conclusions}

In this work we demonstrated that the Lund string model admits a systematic effective theory formulation in which global constraints, such as conservation laws, are implemented through stochastic conditioning rather than ad hoc rejection or endpoint corrections. 
Working in the chiral limit, where the longitudinal-transverse factorization of the Lund kernel becomes exact, we recast hadronization as a Markov process governing the evolution of the string mass $M$. 
As a proxy for satisfying the global constraints on the system, we require  that fragmentation terminate in a specified region $\mathcal{S} = [M_\star, M_{\rm cut}]$, a condition that induces non-Markovian correlations between otherwise independent fragmentation steps. 
Markovianity is restored exactly through the Doob $h$-transform, which renormalizes the bare fragmentation kernel by reweighting each transition according to the success probability $h(M)$.

By exploiting the moment structure of the Lund fragmentation kernel, in which longitudinal moments are order-one while transverse moments scale as inverse powers of $bM^2$, we constructed a tower of effective theories organized by string mass: an ultraviolet fixed point where transport coefficients become scale-invariant, an intermediate running regime where transverse phase space induces controlled $M$-dependence, and a boundary layer near $M_{\rm cut}$ where local diffusion and rare order-one jumps compete. 
This tower structure makes explicit that experimental observables probe fragmentation through three logically distinct stages: local stochastic dynamics governed by the bare fragmentation kernel, non-local renormalization induced by global kinematic constraints, and terminal matching to exclusive hadronic decay. 
By isolating these effects through conditioning and matching, the hadronization process factorizes, allowing the bare UV dynamics to be disentangled from endpoint-induced biases and infrared modeling. 
The analytic composite approximation constructed by matching the boundary and UV solutions captures the qualitative structure of $h(M)$ but retains a residual ${\sim}\,8\%$ discrepancy in the boundary layer, reflecting the structural breakdown of the diffusion approximation where essentially every fragmentation step exits the continuation region. 
This limitation is not an obstacle for practical applications as the exact $h(M)$ can be computed to arbitrary precision by solving the integral backward equation directly, either through matrix inversion or the rapidly convergent Neumann series, both of which retain the full discrete structure of the fragmentation kernel.

This effective theory exhibits genuine Wilsonian renormalization group structure. 
The transport coefficients $\mu(M)$ and $\sigma^2(M)$ are Wilson coefficients that run with mass scale according to $\beta$-functions determined by transverse dynamics, with an anomalous dimension $\gamma(\lambda)$ controlling the flow of the success probability gradient. 
The $\delta$-split scheme reveals a two-scale RG structure in which physical observables are scheme-independent within controlled approximations. 
The success probability $h(M)$, itself a global first-passage observable, serves as the matching quantity that connects the EFT tower. 
Its value and gradient must agree across regime boundaries, determining integration constants and ensuring continuity of physical predictions. 
The gradient of $h(M)$ induces an emergent force $F(M) = \sigma^2(M) \partial_M \ln h(M)$ that biases the evolution toward successful termination, implementing what we term ``budget awareness''. 
This force encodes, in a local and Markovian manner, the effect of global energy-momentum conservation on the stochastic cascade.

This formulation opens several directions for future work. 
Most immediately, it enables a cleaner data-driven extraction of the bare fragmentation function with IR constraint effects systematically removed via the $h$-transform. 
Machine-learning-based hadronization models and methods~\cite{Ilten:2022jfm,Bierlich:2023hzw,Ghosh:2022zdz,Butter:2023vow,Chan:2025flavor} can target the UV kernel directly rather than learning effective modifications that conflate microscopic dynamics with endpoint corrections. 
In neutrino physics, the framework potentially provides a systematic treatment of low-$W$ hadronization where current generators rely on uncontrolled approximations; the boundary-layer yields predictions with no free parameters beyond those fixed at high energy, offering a falsifiable extrapolation to the few-GeV regime critical for oscillation analyses. 
Extending the formalism to gluonic strings is essential for realistic applications; the present $q\bar{q}$ treatment must be generalized to handle the kink dynamics and higher-dimensional state spaces characteristic of multi-parton systems. 
The chiral limit employed throughout this work provides clean analytic control and transparent power counting, but physical applications require systematic inclusion of finite hadron mass effects. 
These enter as relevant operators that explicitly break the longitudinal-transverse factorization, introducing $u$-dependent lower bounds on transverse phase space and mixing UV and IR dynamics. 
The framework developed here provides the baseline against which such corrections can be systematically computed and their parametric scaling established. 
Similarly, the one-dimensional projection onto $M^2 = \Pi^+\Pi^-$ adopted throughout this work does not track the transverse recoil of the remnant string, which accumulates as a random walk under the independent-emission kernel and couples to the termination criterion through $M_{\rm inv}^2 = M^2 - p_{T,\rm rem}^2$. 
Promoting $p_{T,\rm rem}$ to a dynamical variable yields a two-dimensional conditioned process $h(M, p_T)$ in which the Doob transform generates emergent forces in both the mass and transverse directions, biasing the cascade toward lower remnant recoil throughout the evolution rather than only near the termination boundary. 
Further extensions include incorporating flavor and spin degrees of freedom through enlarged state spaces, systematic matching onto cluster decay models at $M_{\rm cut}$, and deploying the hybrid sampling scheme developed here within production Monte Carlo generators. 
More speculatively, the moment structure of the transport coefficients may connect to DGLAP evolution, medium modifications could address heavy-ion applications, and systematic matching to parton showers at the UV scale could enable interleaved shower-hadronization schemes.

The Lund string model has been phenomenologically successful for over four decades, but it has remained largely heuristic, with limited connection to systematic expansion schemes. 
This work demonstrates that the model can be reformulated as an effective theory with well-defined power counting, matching, and renormalization group flow. 
The key observation is that global constraints do not require abandoning local, stochastic dynamics but rather induce a renormalization that can be computed exactly and interpreted physically as an emergent force. 
The resulting factorization of microscopic fragmentation dynamics from infrared constraint effects provides a path toward systematically improvable phenomenology and a transparent connection between microscopic string dynamics and macroscopic hadronic observables.

\section*{Acknowledgements}
A special thank you to the \textsc{MLHAD} collaboration (Beno\^ it Assi, Christian Bierlich, Rikab Gambhir, Phil Ilten, Stephen Mrenna, Manuel Szewc, Michael Wilkinson, and Jure Zupan) as well as Konstantin Matchev for useful discussions and comments on the manuscript. 
This work is supported in part by the Shelby Endowment for Distinguished Faculty at the University of Alabama and by Fermilab via Subcontract 725339.

\newpage
\begin{appendix}

\section{Moment expansion}\label{app:moments}

This appendix derives the exact moment formulas for the longitudinal and transverse fragmentation variables and establishes the power counting used throughout the main text.

\subsection{String mass update and expansion parameter}

From the string mass update derived in \cref{sec:hadronization},
\begin{equation}
  M'^2 = \left( M^2 - \frac{m_\perp^2}{z} \right)(1 - z),
  \label{eq:mass_update_app}
\end{equation}
we factor out $M^2$ and expand:
\begin{equation}\begin{aligned}
  M'^2 
  &= M^2 \left( 1 - \frac{m_\perp^2}{zM^2} \right)(1-z) \\
  &= M^2 \left[ 1 - z - \frac{m_\perp^2}{zM^2} + \frac{m_\perp^2}{M^2} \right] \\
\end{aligned}\end{equation}
Taking the square root and using $1 - 1/z = -(1-z)/z$, we obtain
\begin{equation}
  M' = M \sqrt{1 - z - \frac{m_\perp^2}{M^2}  \left(\frac{1-z}{z}\right)}.
\end{equation}

It is convenient to introduce dimensionless variables
\begin{equation}
  u \equiv z, \quad v \equiv \frac{m_\perp^2}{z M^2}, \quad v \in [0,1],
  \label{eq:uv_def}
\end{equation}
such that $m_\perp^2 = uvM^2$. With these definitions, the string mass update becomes
\begin{equation}
  M' = M \sqrt{(1-u)(1-v)},
  \label{eq:mass_update_uv}
\end{equation}
which is the form used throughout the main text.

For small fractional changes, we define the expansion parameter
\begin{equation}
  \varepsilon \equiv u + v - uv = u + v(1-u),
  \label{eq:epsilon_def}
\end{equation}
such that
\begin{equation}
  M' = M\sqrt{1 - \varepsilon}.
\end{equation}
The mass increment $\Delta M \equiv M' - M$ then admits the Taylor expansion
\begin{equation}\label{eq:DeltaM_expansion}
  \frac{\Delta M}{M} = \sqrt{1-\varepsilon} - 1 = -\frac{1}{2} \varepsilon - \frac{1}{8} \varepsilon^2 - \frac{1}{16} \varepsilon^3 - \frac{5}{128} \varepsilon^4 + \mathcal{O}(\varepsilon^5).
\end{equation}
This expansion should be understood in a statistical sense, organizing expectation values according to powers of $\langle\varepsilon^n\rangle$ rather than as a pointwise approximation valid for every realization.

\subsection{Factorization of the fragmentation kernel}

In the chiral limit adopted throughout this work, the Lund fragmentation kernel defines a joint distribution over $(z, m_\perp^2)$:
\begin{equation}
  f(z, m_\perp^2 \mid M)\,dz\,dm_\perp^2 \propto z^{-1}(1 - z)^a \exp\left( -\frac{b m_\perp^2}{z} \right) dz\,dm_\perp^2,
  \label{eq:lund_ff_app}
\end{equation}
subject to the physical constraints $0 < z < 1$ and $0 \leq m_\perp^2 \leq z M^2$ in the chiral limit.

Under the change of variables in \cref{eq:uv_def}, with $m_\perp^2 = uvM^2$ and Jacobian $dm_\perp^2 = uM^2\,dv$ at fixed $u$, the kernel becomes
\begin{equation}
  f(u,v \mid M)\,du\,dv \propto u^{-1} (1-u)^a e^{-b M^2 v} \cdot uM^2\,du\,dv = (1-u)^a e^{-bM^2 v} M^2\,du\,dv.
\end{equation}
The factor $u^{-1}$ from the original kernel cancels the factor $u$ from the Jacobian, leaving a product of a function of $u$ alone and a function of $v$ alone. 
This establishes the key factorization property: the joint distribution separates as
\begin{equation}
  f(u,v \mid M) = p(u) \, p(v \mid M),
\end{equation}
where the marginal distributions are obtained by normalization.
This factorization is exact in the chiral limit where the lower bound on $v$ becomes scale-independent; finite hadron masses introduce $u$-dependent kinematic constraints that couple the distributions.

Integrating over $v \in [0,1]$ yields the marginal distribution of $u$:
\begin{equation}
  p(u) = (1+a)(1-u)^a, \quad u \in [0,1],
  \label{eq:p_u}
\end{equation}
where the normalization constant is $\mathcal{N}_u = \int_0^1 (1-u)^a du = 1/(1+a)$.

Similarly, integrating over $u \in [0,1]$ yields the marginal distribution of $v$:
\begin{equation}
  p(v \mid M) = \frac{bM^2 e^{-bM^2 v}}{1 - e^{-bM^2}}, \quad v \in [0,1],
  \label{eq:p_v}
\end{equation}
with normalization constant $\mathcal{N}_v = (1 - e^{-bM^2})/(bM^2)$.

The factorization of the joint distribution implies statistical independence:
\begin{equation}
  \mathbb{E}[u^p v^q] = \mathbb{E}[u^p] \cdot \mathbb{E}[v^q],
  \label{eq:moment_factorization}
\end{equation}
where expectations are taken with respect to $p(u)$ and $p(v\,|\,M)$ respectively. This factorization underlies the systematic power counting in the main text.

\subsection{Longitudinal moments}

The moments of $u$ are independent of $M$ and given by beta function integrals:
\begin{equation}\begin{aligned}
  \mathbb{E}[u^p] 
  &= (1+a) \int_0^1 u^p (1-u)^a du \\
  &= (1+a) \frac{\Gamma(p+1)\Gamma(a+1)}{\Gamma(p+a+2)} \\
  &= \frac{(1+a)! \, p!}{(p+a+1)!} \\
  &= \dfrac{p!}{\prod_{k=2}^{p+1}(a+k)}.
\end{aligned}\end{equation}
For the Monash default $a = 0.68$, the first few moments are:
\begin{equation}
  \mathbb{E}[u^p] = 
  \begin{cases}
    \dfrac{1}{a+2} \simeq 0.373 & p = 1, \\[0.1in]
    \dfrac{2}{(a+2)(a+3)} \simeq 0.203 & p = 2, \\[0.1in]
    \dfrac{6}{(a+2)(a+3)(a+4)} \simeq 0.130 & p = 3, \\[0.1in]
    \dfrac{24}{(a+2)(a+3)(a+4)(a+5)} \simeq 0.092 & p = 4.
  \end{cases}
\end{equation}

\subsection{Transverse moments}

The moments of $v$ depend explicitly on $M$ through the dimensionless parameter
\begin{equation}
  \lambda \equiv bM^2,
  \label{eq:lambda_def_app}
\end{equation}
which controls the extent of transverse phase space. Integrating by parts repeatedly yields
\begin{equation}
  \mathbb{E}[v^q] = \int_0^1 v^q p(v \mid M) dv = \dfrac{q!}{\lambda^q (1 - e^{-\lambda})}\left[1 - e^{-\lambda} \sum_{k=0}^{q} \frac{\lambda^k}{k!} \right].
  \label{eq:v_moments_app}
\end{equation}

This expression admits useful limiting forms. For $\lambda \gg 1$ (large $M$):
\begin{equation}
  \mathbb{E}[v^q] \simeq \dfrac{q!}{\lambda^q} + \mathcal{O}(e^{-\lambda}),
  \label{eq:v_moments_UV}
\end{equation}
showing that transverse moments are parametrically suppressed by inverse powers of $M^2$. Conversely, for $\lambda \ll 1$ (small $M$):
\begin{equation}
  \mathbb{E}[v^q] \simeq \dfrac{1}{1+q} - \dfrac{q \, \lambda}{2(q+1)(q+2)} + \mathcal{O}(\lambda^2),
  \label{eq:v_moments_IR}
\end{equation}
which remains order-one as $M \to 0$, indicating the breakdown of the small-$\varepsilon$ expansion near the cutoff.

For $b = 0.98$ GeV$^{-2}$ and representative string masses:
\begin{equation}
  \mathbb{E}[v] \simeq 
  \begin{cases} 
    0.010 & M=10\, \text{GeV}, \\
    0.041 & M=5\, \text{GeV}, \\
    0.420 & M=1\,\text{GeV},
  \end{cases}
  \qquad
  \mathbb{E}[v^2] \simeq 
  \begin{cases} 
    2\times10^{-4} & M=10\, \text{GeV}, \\
    0.003 & M=5\, \text{GeV}, \\
    0.260 & M=1\,\text{GeV}.
  \end{cases}
\end{equation}
These numerical values illustrate the hierarchy $\langle v^q \rangle \ll \langle u^p \rangle$ at large $M$, which underlies the local EFT expansion.

\subsection{Moments of the mass ratio}

For completeness, we record the general formula for moments of $X \equiv (1-u)(1-v)$, which appears in the exact mass update $M' = M\sqrt{X}$. 
Using the binomial theorem and moment factorization:
\begin{equation}\begin{aligned}
  \left\langle X^n \right\rangle 
  &= \left\langle (1-u)^n(1-v)^n \right\rangle \\
  &= \left\langle \sum_{k=0}^n \binom{n}{k} (-u)^k \right\rangle \left\langle \sum_{\ell=0}^n \binom{n}{\ell} (-v)^\ell \right\rangle \\
  &= \sum_{k=0}^n \sum_{\ell=0}^n \binom{n}{k}\binom{n}{\ell} (-1)^{k+\ell} \langle u^k \rangle \langle v^\ell \rangle.
\end{aligned}\end{equation}
Alternatively, expanding $(1-u)(1-v) = 1 - u - v + uv$ and using the multinomial theorem:
\begin{equation}
  \left\langle X^n \right\rangle = \sum_{k=0}^n\sum_{\ell=0}^{n-k} \binom{n}{k}\binom{n-k}{\ell} (-1)^{k+\ell} \langle u^{n-k-\ell} \rangle \langle v^{k+\ell} \rangle.
\end{equation}
These formulas provide the exact cumulants $\kappa_k(M)$ used in the Kramers-Moyal expansion of \cref{subsec:continuum_limit}.

\section{Evaluation of non-local operators}\label{app:tail_operators}

This appendix derives the exact integral representation and asymptotic expansion of the tail operators $\Gamma_{\mathcal{F}}(M)$ and $\Gamma_{\mathcal{S}}(M)$ introduced in \cref{subsec:nonlocal_operators}. 
These operators encode the probability for a single fragmentation step to induce order-one changes in the string mass, directly crossing into the failure region $\mathcal{F} = (0, M_\star)$ or the termination band $\mathcal{S} = [M_\star, M_{\rm cut}]$.

\subsection{Integral representation}

From the definitions in \cref{subsec:nonlocal_operators}, the tail operators are given by
\begin{align}
  \Gamma_{\mathcal{F}}(M)
  &=
  \int_0^1 du \int_0^1 dv\;
  p(u)\,p(v\mid M)\;
  \Theta\!\left(
    \frac{M_\star^2}{M^2}
    -
    (1-u)(1-v)
  \right),
  \label{eq:GammaF_integral}
  \\
  \Gamma_{\mathcal{S}}(M)
  &=
  \int_0^1 du \int_0^1 dv\;
  p(u)\,p(v\mid M)\;
  \Theta\!\left(
    (1-u)(1-v)
    -
    \frac{M_\star^2}{M^2}
  \right)
  \Theta\!\left(
    \frac{M_{\rm cut}^2}{M^2}
    -
    (1-u)(1-v)
  \right),
  \label{eq:GammaT_integral}
\end{align}
where $p(u) = (1+a)(1-u)^a$ and $p(v \mid M) = \lambda e^{-\lambda v}/(1-e^{-\lambda})$ with $\lambda = bM^2$.

The Heaviside functions impose threshold constraints on the allowed values of $(u,v)$. 
For concreteness, consider $\Gamma_{\mathcal{F}}(M)$, which requires $(1-u)(1-v) \leq M_\star^2/M^2$. 
Introducing the dimensionless parameter
\begin{equation}
  \epsilon \equiv \frac{M_{\rm scale}^2}{M^2},
  \label{eq:epsilon_scale}
\end{equation}
the constraint $(1-u)(1-v) \leq \epsilon$ is equivalent to
\begin{equation}
  u \geq 1 - \frac{\epsilon}{1-v} \equiv u_{\text{min}}(v).
  \label{eq:u_threshold}
\end{equation}
This defines a $v$-dependent lower bound on $u$.

\subsection{Exact \texorpdfstring{$u$}{u}-integration}

For fixed $v$, the integral over $u$ can be performed exactly:
\begin{equation}
  \int_{u_{\text{min}}}^1 p(u)\,du 
  = (1+a) \int_{u_{\text{min}}}^1 (1-u)^a du 
  = (1-u_{\text{min}})^{1+a} 
  = \left( \frac{\epsilon}{1-v} \right)^{a+1}.
  \label{eq:u_integral_exact}
\end{equation}
After integrating out $u$, the one-step tail probability for $(1-u)(1-v)\le \epsilon$ may be written as a one-dimensional integral over $v$:
\begin{equation}
  \mathcal{G}(\epsilon)
  =
  \int_0^{1-\epsilon} dv\,p(v\mid M)\left(\frac{\epsilon}{1-v}\right)^{a+1}
  +\int_{1-\epsilon}^{1} dv\,p(v\mid M),
  \label{eq:Gamma_less_1D}
\end{equation}
where $\mathcal{G}(\epsilon)$ denotes the one-step probability for $(1-u)(1-v)\le\epsilon$. The split at $v = 1-\epsilon$ arises because, for $v > 1-\epsilon$, the constraint $(1-u)(1-v) \leq \epsilon$ is automatically satisfied for all $u \in [0,1]$.

For $M\gg M_{\rm scale}$ (i.e., $\epsilon\ll1$), the second term in \cref{eq:Gamma_less_1D} is parametrically small, since it is supported only on the narrow interval $v\in[1-\epsilon,1]$. Isolating this ``saturation'' contribution, we rewrite
\begin{equation}
  \mathcal{G}(\epsilon)
  =
  \epsilon^{a+1}\!\int_0^{1-\epsilon} dv\,p(v\mid M)\,(1-v)^{-(a+1)}
  + \Gamma_{\rm sat}(\epsilon),
  \label{eq:Gamma_less_split}
\end{equation}
where
\begin{equation}
  \Gamma_{\rm sat}(\epsilon)\equiv \int_{1-\epsilon}^{1} dv\,p(v\mid M) = \frac{e^{-\lambda(1-\epsilon)} - e^{-\lambda}}{1 - e^{-\lambda}}.
  \label{eq:Gamma_sat}
\end{equation}
This saturation term is exponentially suppressed for large $\lambda$ and may be treated as a higher-order correction in the overlap region with the local EFTs.

\subsection{Closed-form expression}

To evaluate the first term in \cref{eq:Gamma_less_split}, we substitute $t = 1-v$ to obtain
\begin{equation}
  \int_0^{1-\epsilon} dv\,p(v\mid M)\left(\frac{\epsilon}{1-v}\right)^{1+a}
  =
  \epsilon^{1+a}\frac{\lambda e^{-\lambda}}{1-e^{-\lambda}}
  \int_{\epsilon}^{1} dt\,t^{-(1+a)}e^{\lambda t}.
  \label{eq:integral_t}
\end{equation}
The integral over $t$ is expressed in terms of the incomplete gamma function. Using the identity
\begin{equation}
  \int dt\,t^{-(a+1)}e^{\lambda t}
  =
  -(-\lambda)^a\,\Gamma\!\big(-a,-\lambda t\big) + \text{const},
  \label{eq:incomplete_gamma}
\end{equation}
where $\Gamma(s,x) = \int_x^\infty t^{s-1} e^{-t} dt$ is the upper incomplete gamma function, we find
\begin{equation}
  \int_{\epsilon}^{1} dt\,t^{-(a+1)}e^{\lambda t}
  =
  -(-\lambda)^a \left[ \Gamma\!\big(-a,-\lambda\big) - \Gamma\!\big(-a,-\lambda\epsilon\big) \right].
\end{equation}
Substituting into \cref{eq:integral_t} and combining with the saturation term yields the exact closed form
\begin{equation}
  \mathcal{G}(\epsilon)
  =
  \epsilon^{a+1}\frac{e^{-\lambda}}{1-e^{-\lambda}}
  (-\lambda)^{a+1}\Big[\Gamma\!\big(-a,-\lambda\epsilon\big)-\Gamma\!\big(-a,-\lambda\big)\Big]
  +\frac{e^{-\lambda(1-\epsilon)}-e^{-\lambda}}{1-e^{-\lambda}}.
  \label{eq:Gamma_less_closed}
\end{equation}
An equivalent representation uses the generalized exponential integral $E_\nu(z) = \int_1^\infty t^{-\nu} e^{-zt}\, dt$, which is related to the incomplete gamma function by
\begin{equation}
  E_\nu(z) = z^{\nu-1} \Gamma(1-\nu, z),
  \qquad \text{or equivalently} \qquad
  \Gamma(-a, z) = z^{-a} E_{a+1}(z).
  \label{eq:E_Gamma_relation}
\end{equation}
Substituting this relation into \cref{eq:Gamma_less_closed} yields
\begin{equation}
  \mathcal{G}(\epsilon; M) =
  \frac{\lambda}{e^{\lambda}-1}
  \Big[\epsilon^{a+1} E_{a+1}(-\lambda) - \epsilon \, E_{a+1}(-\lambda\epsilon)\Big]
  +
  \frac{e^{\lambda\epsilon}-1}{e^{\lambda}-1},
  \label{eq:Gamma_less_E}
\end{equation}
which is the form quoted in the main text.
The failure and termination-band operators are then obtained by evaluating $\mathcal{G}$ at the appropriate scales:
\begin{equation}
  \Gamma_{\mathcal{F}}(M)=\mathcal{G}(\epsilon_\star),\qquad
  \Gamma_{\mathcal{S}}(M)=\mathcal{G}(\epsilon_{\rm cut})-\mathcal{G}(\epsilon_\star),
  \label{eq:tail_from_Gamma_less}
\end{equation}
where
\begin{equation}
  \epsilon_\star\equiv \frac{M_\star^2}{M^2},\qquad
  \epsilon_{\rm cut}\equiv \frac{M_{\rm cut}^2}{M^2}.
\end{equation}

\subsection{Moment expansion in the transverse variable}

To make contact with the local EFT and establish power counting, we expand the integrand in \cref{eq:Gamma_less_split} in powers of $v$. 
The factor $(1-v)^{-(a+1)}$ appearing in the first term admits the absolutely convergent binomial series for $v \in [0,1)$:
\begin{equation}
  (1-v)^{-(a+1)}
  =
  \sum_{n=0}^\infty \frac{(a+1)_n}{n!}\,v^n,
  \label{eq:binom_expansion}
\end{equation}
where
\begin{equation}
  (a+1)_n \equiv \frac{\Gamma(a+1+n)}{\Gamma(a+1)} = (a+1)(a+2)\cdots(a+n)
  \label{eq:Pochhammer}
\end{equation}
is the rising Pochhammer symbol. 
This identity follows from the generalized binomial theorem:
\begin{equation}
  (1-v)^{-\alpha} = \sum_{n=0}^\infty \binom{\alpha + n - 1}{n} v^n = \sum_{n=0}^\infty \frac{(\alpha)_n}{n!} v^n.
\end{equation}

Substituting \cref{eq:binom_expansion} into \cref{eq:Gamma_less_split} and exchanging the sum and integral yields
\begin{equation}
  \mathcal{G}(\epsilon)
  =
  \epsilon^{a+1}\sum_{n=0}^\infty \frac{(a+1)_n}{n!}\,\langle v^n\rangle_\lambda
  \;+\;
  \Gamma_{\rm sat}(\epsilon)
  \;+\;{\cal O}\!\left(\epsilon^{a+2}\right),
  \label{eq:Gamma_moment_series}
\end{equation}
where the transverse moments are
\begin{equation}
  \langle v^n\rangle_\lambda
  \equiv
  \int_0^1 dv\,p(v\mid M)\,v^n = \int_0^1 dv\, \frac{\lambda e^{-\lambda v}}{1-e^{-\lambda}} v^n,
  \label{eq:v_moments_def}
\end{equation}
as computed explicitly in appendix~\ref{app:moments}. 
The remainder ${\cal O}(\epsilon^{a+2})$ arises from extending the upper limit $1-\epsilon\to1$ in the first term of \cref{eq:Gamma_less_split}; this correction is parametrically subleading relative to the leading $\epsilon^{a+1}$ scaling.

Defining the expansion coefficients
\begin{equation}
  \alpha_n = \frac{(a+1)_n}{n!} = \frac{(a+1)(a+2)\cdots(a+n)}{n!},
  \label{eq:alpha_n_def}
\end{equation}
we obtain the systematic tail-operator series
\begin{equation}
  \Gamma_{\mathcal{F}}(M)
  =
  \left(\frac{M_\star}{M}\right)^{2(a+1)}
  \sum_{n=0}^\infty \alpha_n\,\langle v^n\rangle_\lambda
  \;+\;
  \Gamma_{\rm sat}(\epsilon_\star)
  \;+\;{\cal O}\!\left(\frac{M_\star^{2(a+2)}}{M^{2(a+2)}}\right),
  \label{eq:GammaF_series}
\end{equation}
and similarly
\begin{equation}
  \Gamma_{\mathcal{S}}(M)
  =
  \left[
    \left(\frac{M_{\rm cut}}{M}\right)^{2(a+1)}
    -
    \left(\frac{M_\star}{M}\right)^{2(a+1)}
  \right]
  \sum_{n=0}^\infty \alpha_n\,\langle v^n\rangle_\lambda
  \;+\;
  \Gamma_{\rm sat}(\epsilon_{\rm cut})-\Gamma_{\rm sat}(\epsilon_\star)
  \;+\;\cdots.
  \label{eq:GammaT_series}
\end{equation}

\subsection{Physical interpretation and power counting}

Equations \cref{eq:GammaF_series,eq:GammaT_series} exhibit the parametric structure of the tail operators explicitly. Each contribution scales as $(M_{\rm scale}/M)^{2(a+1)}$, which is strongly suppressed for $M \gg M_{\rm scale}$. 
The sum over $n$ represents a systematic expansion in transverse moments $\langle v^n \rangle_\lambda \sim \lambda^{-n}$, with each higher-$n$ term contributing an additional suppression factor $\lambda^{-1} = (bM^2)^{-1}$.

This double hierarchy---power suppression in $(M_{\rm scale}/M)$ and moment suppression in $\lambda^{-1}$---establishes the parametric ordering of tail effects relative to local drift-diffusion transport. 
In the UV regime where both $M \gg M_{\rm scale}$ and $\lambda \gg 1$, the tail operators are exponentially small and the local EFT provides a complete description. 
In the boundary layer where $M \sim M_{\rm cut}$ and $\lambda \sim \mathcal{O}(1)$, the tail operators become order-one and must be retained explicitly, as discussed in \cref{sec:h_global}.

The series in \cref{eq:GammaF_series,eq:GammaT_series} is not an expansion of the exact closed-form expression \cref{eq:Gamma_less_closed}, but rather a systematic organization of the endpoint contributions in the overlap region with the local EFT. 
Truncation at fixed $n$ defines a tail EFT with controlled accuracy, while resumming the full series (or using the exact incomplete gamma function representation) provides the complete non-local structure.

\subsection{Numerical evaluation}

In practice, we evaluate the tail operators using one of three methods depending on the regime:
\begin{enumerate}
\item \textbf{Large $M$ ($\epsilon \ll 1$, $\lambda \gg 1$)}: Use the moment expansion \cref{eq:GammaF_series,eq:GammaT_series} truncated at low order. The saturation terms are exponentially suppressed and can be dropped. 
This regime corresponds to the UV and running local EFTs.

\item \textbf{Intermediate $M$}: Use the exact closed form \cref{eq:Gamma_less_closed} evaluated numerically. This provides a faithful interpolation across all mass scales without approximation.

\item \textbf{Near-boundary ($M \sim M_{\rm cut}$)}: Evaluate the full integral \cref{eq:Gamma_less_1D} numerically, or use the exact closed form. In this regime, both the saturation term and the leading power-law contribution are order-one.
\end{enumerate}

The analytic expressions derived here enable efficient computation of the success probability $h(M)$ through the backward equation solved in \cref{sec:h_global}, and they provide the matching data required for the renormalized drift in \cref{sec:doob_h}.

\section{Constrained Markovian dynamics}
\label{app:constrained_dynamics}

This appendix develops a pedagogical toy model that illustrates the key concepts of trajectory conditioning and Doob $h$-transforms in a simpler setting than string hadronization. 
We consider Brownian motion with drift subject to endpoint constraints, deriving the renormalized dynamics explicitly and interpreting the results physically. 
This example builds intuition for the stochastic analysis techniques employed in the main text while avoiding the technical complications of non-local operators and multi-scale matching.

The analogy to hadronization is direct: the Brownian particle corresponds to the evolving string mass, the drift represents the bare Lund dynamics, and the endpoint constraint mimics the termination requirement $M \in \mathcal{S}$. 
The $h$-function here plays the same role as the success probability in \cref{sec:h_global}, and the emergent force has the same mathematical structure as the conditioning-induced drift in \cref{sec:doob_h}.

\subsection{Brownian motion with drift}

\subsubsection{Discrete random walk}

Consider a discrete Gaussian random walk starting at the origin:
\begin{equation}
  W_{n+1} = W_n + \sqrt{\Delta t} \,\, \xi_n, \quad \xi_n \sim \mathcal{N}(0,1), \quad W_0 = 0,
  \label{eq:discrete_walk}
\end{equation}
where $\xi_n$ are independent standard normal random variables and $\Delta t$ is the time step. After $n = t / \Delta t$ steps, the position is
\begin{equation}
  W_n = \sum_{k=0}^{n-1} \sqrt{\Delta t} \,\, \xi_k,
\end{equation}
with mean and variance
\begin{equation}
  \mathbb{E}[ W_n ] = 0, \qquad \text{Var}(W_n) = n \Delta t = t.
\end{equation}

In the continuum limit $\Delta t \to 0$ and $n \to \infty$ with $n\Delta t = t$ held fixed, this discrete process converges to a \emph{Wiener process} (or standard Brownian motion) $W_t$. The Wiener process is characterized by continuous but nowhere-differentiable paths, with independent Gaussian increments:
\begin{equation}
  W_t - W_s \sim \mathcal{N}(0, t-s) \quad \text{for } t > s.
\end{equation}

\subsubsection{Adding drift and diffusion}

Now consider a more general stochastic process $S_t$ evolving in discrete time as
\begin{equation}
  S_{n+1} = S_n + b\, \Delta t + \sigma \sqrt{\Delta t} \,\, \xi_n,
  \label{eq:discrete_drift_diffusion}
\end{equation} 
where $b$ is a constant drift and $\sigma$ controls the amplitude of stochastic fluctuations. In the continuum limit, this becomes a stochastic differential equation (SDE):
\begin{equation}\label{eq:arithmetic_brownian_SDE}
  d S_t = b\, dt + \sigma\, dW_t, \qquad S_0 = s.
\end{equation}
Here $dW_t$ represents an infinitesimal increment of the Wiener process, with $\mathbb{E}[dW_t] = 0$ and $\text{Var}[dW_t] = dt$.

Integrating both sides with respect to time gives the exact solution
\begin{equation}
  S_t = s + bt + \sigma W_t,
  \label{eq:brownian_solution}
\end{equation}
from which we immediately read off
\begin{equation}
  \mathbb{E}[S_t] = s + b\,t, \qquad \text{Var}(S_t) = \sigma^2 t, \qquad S_t \sim \mathcal{N}(s + bt, \sigma^2 t).
  \label{eq:brownian_statistics}
\end{equation}
The process drifts deterministically at rate $b$ while diffusing stochastically with variance growing linearly in time.

\subsubsection{It\^o's lemma and the generator}

To understand how smooth functions of $S_t$ evolve under the stochastic dynamics, we need It\^o's lemma. For a twice-differentiable function $f(S_t,t)$, the differential $df$ is given by
\begin{equation}\label{eq:itos_lemma}
  df = \left(\partial_t f + b\, \partial_S f + \frac{1}{2}\sigma^2 \partial^2_{S} f \right) dt + \sigma \, \partial_S f \, dW_t,
\end{equation}
where $\partial_S \equiv \partial/\partial S$ and $\partial^2_S \equiv \partial^2/\partial S^2$. The deterministic part (coefficient of $dt$) defines the \emph{generator} of the stochastic process:
\begin{equation}
  \mathcal{L} \equiv b\, \partial_S + \frac{1}{2}\sigma^2 \partial^2_{S}.
  \label{eq:brownian_generator_def}
\end{equation}

The generator controls the evolution of expectation values. Taking expectations on both sides of \cref{eq:itos_lemma} and using $\mathbb{E}[dW_t] = 0$, we find
\begin{equation}
  \frac{d}{dt} \mathbb{E}[f(S_t,t)] = \mathbb{E}\left[(\partial_t + \mathcal{L})f\right].
  \label{eq:expectation_evolution}
\end{equation}
The second-derivative term $\frac{1}{2}\sigma^2 \partial^2_S f$ is the signature of diffusion: it captures how spreading of the probability distribution affects the mean of $f$. This term arises from the quadratic variation of the Wiener process and has no analogue in ordinary calculus.

\subsection{Constrained Brownian motion}

\subsubsection{Martingales and conserved charges}

A particularly important class of observables consists of functions whose expectation values are \emph{constant in time}. From \cref{eq:expectation_evolution}, this requires
\begin{equation}
  (\partial_t + \mathcal{L})f = 0.
  \label{eq:martingale_condition}
\end{equation}
Such a function defines a \emph{martingale}: a quantity that, although fluctuating along individual trajectories, has zero systematic drift when averaged over the ensemble. 
This is the stochastic counterpart of a conserved charge in deterministic dynamics, representing information that is transported through time without loss or gain on average.

For time-independent functions satisfying $\mathcal{L} f = 0$, the martingale property reduces to invariance under the generator alone. 
These functions are called \emph{harmonic} and play a central role in what follows.

\subsubsection{The backward equation}

A particularly useful class of martingales encodes conditional expectations of future observables. 
Consider a terminal observable $\mathcal{O}(S_T)$ measured at time $T$, and define
\begin{equation}
  u(S,t) = \mathbb{E}\left[ \mathcal{O}(S_T) \,\big|\, S_t = S \right],
  \label{eq:u_def}
\end{equation}
the expected value of $\mathcal{O}$ given that the system is at position $S$ at time $t < T$. The boundary condition is
\begin{equation}
  u(S, T) = \mathcal{O}(S).
\end{equation}

For such functions to be martingales, they must satisfy
\begin{equation}\label{eq:backward_evolution}
  \partial_t u + b\, \partial_S u + \frac{1}{2}\sigma^2 \partial^2_{S} u = 0.
\end{equation}
This is the \emph{Kolmogorov backward equation}, which propagates the terminal condition $\mathcal{O}(S)$ backward in time to produce the conditional expectation field $u(S,t)$ throughout the domain. 
The name ``backward'' emphasizes that we solve from $T$ toward $t=0$, in contrast to the forward Fokker-Planck equation that evolves probability densities forward in time.

\subsubsection{Example: first-passage probability}

As a concrete example, consider the probability that a trajectory starting from $S$ at time $t$ will end inside a target region $\mathcal{R} \equiv [\alpha, \beta]$ at time $T$:
\begin{equation}
  u(S,t) = \mathbb{P}(S_T \in \mathcal{R} \,\big|\, S_t = S),
  \label{eq:success_probability_toy}
\end{equation}
with boundary condition
\begin{equation}
  u(S,T) = 
  \begin{cases} 
    1, & S \in \mathcal{R}, \\
    0, & S \notin \mathcal{R}.
  \end{cases}
\end{equation}
This is the direct analogue of the success probability $h(M)$ in the hadronization problem.

\subsubsection{Solution via heat equation}

To solve \cref{eq:backward_evolution}, we eliminate the drift by changing variables. 
Define
\begin{equation}
  x \equiv S + b(T - t),
  \label{eq:moving_frame}
\end{equation}
which represents position in a frame moving with the drift. Substituting $v(x,t) = u(S,t)$ into \cref{eq:backward_evolution} transforms it into the standard heat equation:
\begin{equation}
  \partial_t v + \frac{1}{2} \sigma^2 \partial_x^2 v = 0.
  \label{eq:heat_equation}
\end{equation}
The general solution is a Gaussian convolution with the terminal data:
\begin{equation}
  v(x,t) = \int_{\mathcal{R}} \frac{1}{\sqrt{2\pi \sigma^2(T - t)}} \exp \left[ -\frac{(y-x)^2}{2\sigma^2 (T - t)} \right] dy.
\end{equation}

For the interval $\mathcal{R} = [\alpha, \beta]$, the integral evaluates to a difference of cumulative distribution functions. 
Transforming back to the original variables $(S,t)$ gives
\begin{equation}\label{eq:toy_u_solution}
  u(S,t) = \Phi\left( \frac{\beta - S - b (T - t)}{\sigma \sqrt{T - t}}  \right) - \Phi\left( \frac{\alpha - S - b (T - t)}{\sigma \sqrt{T - t}} \right),
\end{equation}
where $\Phi(z) = \int_{-\infty}^z \frac{1}{\sqrt{2\pi}} e^{-s^2/2} ds$ is the standard normal cumulative distribution function.

This solution encodes the complete global probability field: for any starting point $(S,t)$, it gives the probability of success at time $T$. 
The field ``knows'' about all possible future trajectories and their outcomes.

\begin{figure}[t]
\centering
\includegraphics[width=0.6\textwidth]{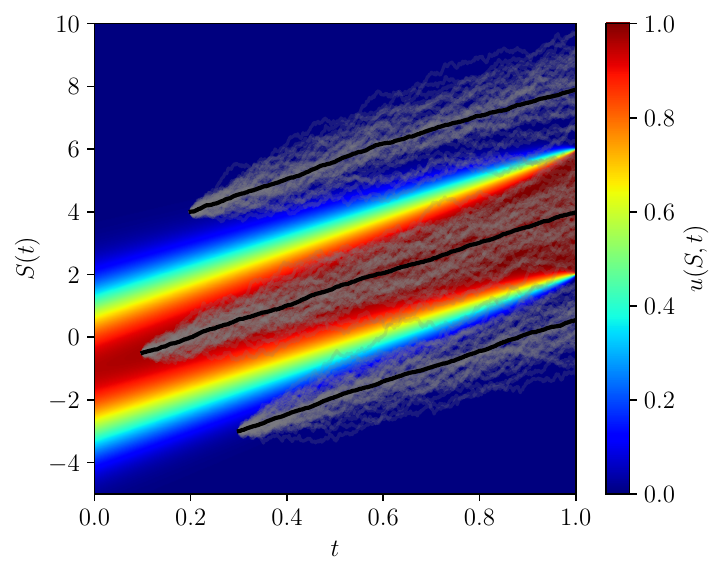}
\caption{Success probability field $u(S,t)$ for Brownian motion with drift $b=5$ and diffusion $\sigma=1$, with sample trajectories overlaid. The field interpolates from $u=0$ (certain failure) far from the target region $\mathcal{R} = [\alpha, \beta] = [2, 6]$ to $u=1$ (certain success) deep inside $\mathcal{R}$.}
\label{fig:u_field}
\end{figure}

\subsection{Doob \texorpdfstring{$h$}{h}-transform and renormalized dynamics}

\subsubsection{The filtering problem}

In physical applications, we often wish to impose global constraints on trajectories. 
For instance, we may require that all simulated histories end in the target region $\mathcal{R}$, analogous to imposing energy-momentum conservation in hadronization. 
Naively, this is implemented by \emph{filtering}: run the bare stochastic process and discard trajectories that fail the constraint.

Filtering is conceptually simple but computationally inefficient. 
If the success probability is small, most trajectories are wasted. Moreover, filtering obscures the relationship between the bare dynamics and the constrained ensemble: the act of discarding trajectories introduces apparent non-Markovian correlations.

\subsubsection{The Doob \texorpdfstring{$h$}{h}-transform}

The Doob $h$-transform provides an elegant alternative. 
Instead of filtering after the fact, we \emph{reweight} the transition probabilities during evolution to favor configurations that will eventually succeed. 
The key insight is that the success probability $u(S,t)$ itself serves as the reweighting factor.

Define $h(S,t) \equiv u(S,t)$ as our renormalization function. 
The $h$-transformed generator is
\begin{equation}
  \mathcal{L}_h \equiv h^{-1} \mathcal{L} h.
  \label{eq:h_transform_def}
\end{equation}
Expanding this explicitly:
\begin{align}
  \mathcal{L}_h f 
  &= h^{-1} \left[ b \, \partial_S(hf) + \tfrac{1}{2}\sigma^2 \partial^2_S(hf) \right] \nonumber \\
  &= h^{-1} \left[ b h \partial_S f + b f \partial_S h + \tfrac{1}{2}\sigma^2 h \partial^2_S f + \sigma^2 (\partial_S h)(\partial_S f) + \tfrac{1}{2}\sigma^2 f \partial^2_S h \right] \nonumber \\
  &= b \partial_S f + \sigma^2 h^{-1} (\partial_S h) \partial_S f + \tfrac{1}{2}\sigma^2 \partial^2_S f + \underbrace{b h^{-1} \partial_S h + \tfrac{1}{2}\sigma^2 h^{-1} \partial^2_S h}_{=0} \, f.
\end{align}
The terms proportional to $f$ vanish because $h$ satisfies the backward equation $\mathcal{L} h = 0$. 
Combining terms, the transformed generator takes the form
\begin{equation}
  \mathcal{L}_h = b_{\rm eff} \, \partial_S + \tfrac{1}{2}\sigma^2 \partial^2_S,
  \label{eq:transformed_generator}
\end{equation}
where the \emph{renormalized drift} is
\begin{equation}
  b_{\rm eff}(S,t) = b + \sigma^2 \partial_S \ln h(S,t).
  \label{eq:renormalized_drift_toy}
\end{equation}

The diffusion coefficient $\sigma^2$ is unchanged, but the drift acquires an additional term proportional to the gradient of $\ln h$. 
This extra drift acts as an emergent force pulling trajectories toward regions of high success probability, exactly analogous to the force $F(M) = \sigma^2(M) \partial_M \ln h(M)$ in hadronization.

\subsubsection{Explicit formula}

\begin{figure}[t]
\centering
\includegraphics[width=0.49\textwidth]{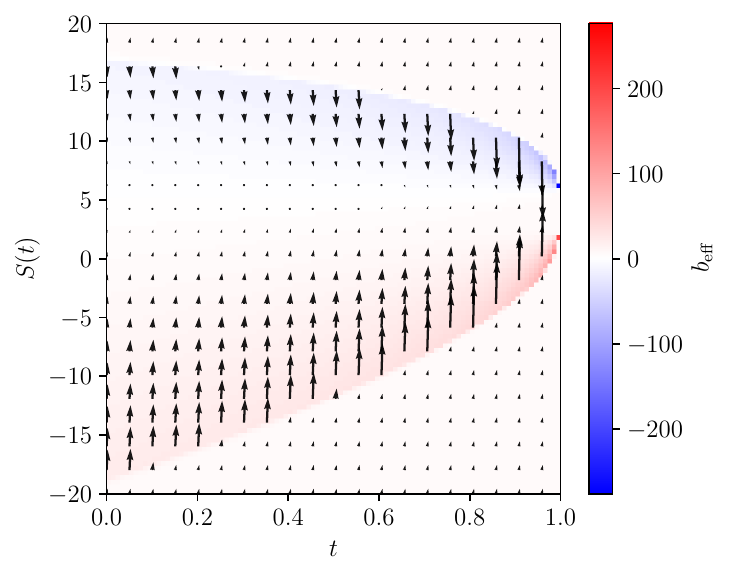}
\includegraphics[width=0.49\textwidth]{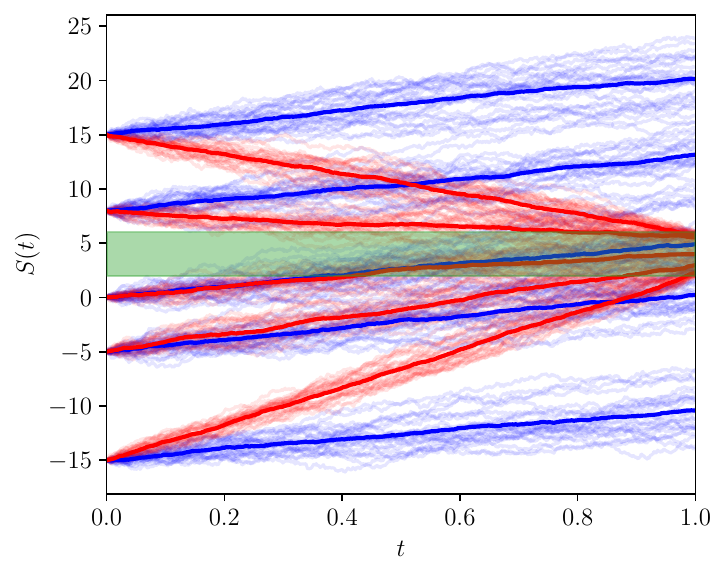}
\caption{\textbf{Left:} Renormalized drift $b_{\rm eff}(S,t)$ for the toy model with $b=5$, $\sigma=1$, and target region $\mathcal{R} = [2, 6]$. The drift is reduced below $\alpha=2$ (pulling trajectories toward $\mathcal{R}$) and enhanced above $\beta=6$ (slowing escape). \textbf{Right:} Sample trajectories comparing bare (dashed) and $h$-transformed (solid) dynamics. Conditioned trajectories are biased to remain within $\mathcal{R}$.}
\label{fig:beff_field}
\end{figure}

For the solution in \cref{eq:toy_u_solution}, the logarithmic derivative is
\begin{equation}
  \partial_S \ln h = \frac{1}{h(S,t)} \, \partial_S h = \frac{\phi\left( \dfrac{\alpha - S - b (T - t)}{\sigma \sqrt{T - t}}  \right) - \phi\left( \dfrac{\beta - S - b (T - t)}{\sigma \sqrt{T - t}}  \right)}{\sigma h(S,t) \sqrt{T - t}},
  \label{eq:d_ln_h_explicit}
\end{equation}
where $\phi(z) = \frac{1}{\sqrt{2\pi}} e^{-z^2/2}$ is the standard normal probability density function. 
Substituting into \cref{eq:renormalized_drift_toy} gives the effective drift field $b_{\rm eff}(S,t)$ shown in Figure~\ref{fig:beff_field}.

\subsection{Effective theory in different regimes}

The renormalized drift \cref{eq:renormalized_drift_toy} interpolates between qualitatively different behaviors depending on the system's configuration. 
To organize these regimes, define:
\begin{itemize}
\item $\Delta \equiv T - t$: remaining time until terminal evaluation,
\item $x \equiv S + b\Delta$: expected position at time $T$ under bare drift,
\item $\varepsilon \equiv \sigma \sqrt{\Delta}$: characteristic diffusion spread over remaining time,
\item $d \equiv \min(\beta - x, \, x - \alpha)$: distance from expected endpoint to nearest boundary.
\end{itemize}
These quantities define dimensionless ratios that control the structure of the effective theory, in direct analogy to the EFT power counting in hadronization.

\subsubsection{Regime 1: Deep inside \texorpdfstring{$\mathcal{R}$}{R} (\texorpdfstring{$d \gg \varepsilon$}{d >> varepsilon})}

When the expected endpoint $x = S + b\Delta$ lies well within the target region, success is essentially guaranteed and the success probability approaches unity:
\begin{equation}
  h(S,t) \simeq 1 - \frac{\varepsilon}{d \sqrt{2\pi}} \exp\left[-\frac{d^2}{2\varepsilon^2}\right] = 1 - \mathcal{O}\left(e^{-d^2/2\varepsilon^2}\right).
\end{equation}
The correction is exponentially suppressed in the ratio $(d/\varepsilon)^2$, indicating that diffusive fluctuations are unlikely to push the trajectory outside $\mathcal{R}$. Differentiating:
\begin{equation}
  \partial_S \ln h = -\frac{1}{h} \frac{\partial h}{\partial S} \simeq \mathcal{O}\left(\frac{d}{\varepsilon^2} e^{-d^2/2\varepsilon^2}\right),
\end{equation}
which is also exponentially small. Consequently, the renormalized drift reduces to
\begin{equation}
  b_{\rm eff} \simeq b + \mathcal{O}\left( e^{-d^2/2\varepsilon^2} \right),
\end{equation}
recovering the bare dynamics. This is the analogue of the UV regime in hadronization, where the success probability $h(M) \approx h_\infty$ is approximately constant and the emergent force vanishes.

\subsubsection{Regime 2: Far outside \texorpdfstring{$\mathcal{R}$} (\texorpdfstring{$x - \beta \gg \varepsilon$}{x-beta >> varepsilon} or \texorpdfstring{$\alpha - x \gg \varepsilon$}{alpha-x >> varepsilon})}

Consider a trajectory whose expected endpoint lies far above the target region: $x - \beta \gg \varepsilon$. In this regime, success requires a large downward fluctuation, and the success probability is small:
\begin{equation}
  h(S,t) \simeq \Phi\left( \frac{\beta - x}{\varepsilon} \right) \simeq \phi\left( \frac{\beta - x}{\varepsilon} \right) \frac{\varepsilon}{\beta - x},
\end{equation}
where the second approximation uses the tail asymptotics of the normal CDF: $\Phi(-z) \approx \phi(z)/z$ for $z \gg 1$. Taking the logarithmic derivative:
\begin{equation}
  \partial_S \ln h \simeq \frac{\beta - x}{\varepsilon^2},
\end{equation}
yielding the renormalized drift
\begin{equation}
  b_{\rm eff} \simeq b + \sigma^2 \frac{\beta - x}{\varepsilon^2} = b + \frac{\beta - S - b\Delta}{\Delta} = \frac{\beta - S}{\Delta}.
  \label{eq:bridge_drift}
\end{equation}

This is a \emph{Brownian bridge} to the upper boundary: the effective drift points directly toward $\beta$ with strength inversely proportional to the remaining time. 
The system ``knows'' it must reach $\beta$ to succeed and adjusts its dynamics accordingly. 
Similarly, trajectories far below $\mathcal{R}$ experience a bridge toward $\alpha$.
This regime is analogous to the boundary layer in hadronization, where the emergent force becomes order-one and dominates the bare drift.

\subsubsection{Regime 3: Edge of \texorpdfstring{$\mathcal{R}$}{R} (\texorpdfstring{$d \sim \varepsilon$}{d ~ varepsilon})}

When the expected endpoint lies near a boundary of $\mathcal{R}$, diffusive fluctuations are order-one relative to the distance to the edge, and the success probability exhibits rapid spatial variation. 
Consider $x \approx \alpha$ with $d = x - \alpha \sim \varepsilon$. The success probability is
\begin{equation}
  h(S,t) \simeq \Phi\left( \frac{d}{\varepsilon} \right) \simeq \frac{1}{2} + \frac{1}{\sqrt{2\pi}} \frac{d}{\varepsilon} + \mathcal{O}\left(\frac{d^3}{\varepsilon^3}\right),
\end{equation}
where we expanded around $d=0$. 
The logarithmic derivative is
\begin{equation}
  \partial_S \ln h \simeq \frac{2}{\varepsilon \sqrt{2\pi}} \cdot \frac{1}{1 + \sqrt{2/\pi} \, d/\varepsilon} \simeq \frac{\sqrt{2/\pi}}{\varepsilon} \left(1 - \frac{\sqrt{2/\pi} d}{\varepsilon} + \cdots \right),
\end{equation}
and the renormalized drift becomes
\begin{equation}
  b_{\rm eff} \simeq b + \frac{\sigma}{\sqrt{2\pi\Delta}} - \frac{d}{\pi \Delta} + \cdots.
\end{equation}

The leading correction is $\mathcal{O}(\sigma/\sqrt{\Delta})$, intermediate between the exponentially suppressed interior regime and the $\mathcal{O}(1/\Delta)$ bridge regime. 
This boundary layer exhibits strong sensitivity to position: small changes in $S$ produce order-one changes in the renormalized drift, pulling marginal trajectories toward safety or pushing hopeless ones toward failure.

\subsubsection{Regime 4: Narrow window (\texorpdfstring{$\beta - \alpha \ll \varepsilon$}{beta-alpha << varepsilon})}

If the target region $\mathcal{R}$ is narrow compared to the diffusion scale---$\beta - \alpha \ll \sigma\sqrt{\Delta}$---then reaching $\mathcal{R}$ requires precise control. 
In this limit, the success probability becomes exponentially sensitive to position:
\begin{equation}
  h(S,t) \simeq \frac{\beta - \alpha}{\varepsilon \sqrt{2\pi}} \exp\left[ -\frac{(x - \bar{x})^2}{2\varepsilon^2} \right],
\end{equation}
where $\bar{x} = (\alpha + \beta)/2$ is the center of $\mathcal{R}$. 
This is approximately a Gaussian profile centered at $x = \bar{x}$, with width $\varepsilon$. 
The logarithmic derivative is
\begin{equation}
  \partial_S \ln h \simeq -\frac{x - \bar{x}}{\varepsilon^2},
\end{equation}
yielding
\begin{equation}
  b_{\rm eff} \simeq b - \sigma^2 \frac{x - \bar{x}}{\varepsilon^2} = \frac{\bar{x} - S}{\Delta}.
\end{equation}

The renormalized drift points toward the \emph{center} of the narrow window, creating a harmonic-oscillator-like potential well. 
Trajectories are pulled toward $\bar{x}$ with restoring force proportional to displacement, analogous to how boundary conditions in hadronization bias the cascade toward the center of the termination band when $M_{\rm cut} - M_\star$ is small.

\subsubsection{Regime 5: Closing horizon (\texorpdfstring{$\Delta \to 0$}{Delta -> 0})}

As the terminal time $T$ approaches, the remaining diffusion becomes negligible: $\varepsilon = \sigma\sqrt{\Delta} \to 0$. 
In this limit, the success probability reduces to the step function:
\begin{equation}
  \lim_{\Delta \to 0} h(S,t) = 
  \begin{cases}
    1 & \text{if } S \in \mathcal{R}, \\
    0 & \text{if } S \notin \mathcal{R}.
  \end{cases}
\end{equation}

For $S$ already inside $\mathcal{R}$, the trajectory has succeeded and $h=1$, so $\partial_S \ln h = 0$ and $b_{\rm eff} = b$: no correction is needed. 
For $S$ outside $\mathcal{R}$, the trajectory has failed ($h \approx 0$) and the renormalized drift formally diverges, reflecting the impossibility of reaching $\mathcal{R}$ in vanishing time.

More precisely, in the transition layer of width $\varepsilon$ near the boundaries, the drift \cref{eq:bridge_drift} scales as $1/\Delta$, becoming arbitrarily strong. This ``last-ditch'' effort to push trajectories into $\mathcal{R}$ mirrors the behavior of the emergent force in hadronization as the cascade approaches $M_{\rm cut}$ with insufficient mass remaining.

\subsection{Connection to hadronization}

The Brownian motion example exhibits all the key features of the hadronization problem:

\begin{enumerate}
\item \textbf{Global constraint encoded locally}: The terminal condition $S_T \in \mathcal{R}$ is a non-local constraint on the full trajectory, yet its effect enters the effective dynamics through the \emph{local} function $h(S,t)$ and its gradient.

\item \textbf{Doob transform preserves Markovianity}: Although conditioning on success appears to introduce trajectory-level correlations, the $h$-transform restores a local, Markovian description with modified transport coefficients.

\item \textbf{Emergent force}: The additional drift $\sigma^2 \partial_S \ln h$ acts as an effective force that ``pulls'' trajectories toward configurations with higher success probability, analogous to the force $\sigma^2(M) \partial_M \ln h(M)$ in hadronization.

\item \textbf{Regime-dependent behavior}: The effective theory exhibits qualitatively different behaviors depending on dimensionless ratios (here $d/\varepsilon$, in hadronization $(M/M_{\rm cut})$ and $\lambda = bM^2$), justifying the EFT tower structure.

\item \textbf{Backward equation as central tool}: The success probability $h$ satisfies a backward differential equation that propagates boundary conditions into the interior, exactly paralleling the role of $h(M)$ in Section~\ref{sec:h_global}.
\end{enumerate}

The analogy is not merely qualitative. 
The mathematical structure---similarity transformation of the generator, renormalization of drift but not diffusion, emergence of a force proportional to $\partial \ln h$---is identical. 
The complications in hadronization arise from the nontrivial functional form of $\mu(M)$ and $\sigma^2(M)$, the presence of non-local tail operators, and the need for multi-scale matching. 
But the conceptual framework developed here carries over directly, making this toy model a valuable pedagogical tool for understanding the full theory.

\end{appendix}

\newpage
\bibliography{ref}{}
\bibliographystyle{JHEP}

\end{document}